# STAR NRE: solving supernova selection effects with set-based truncated auto-regressive neural ratio estimation

Konstantin Karchev[a] and Roberto Trotta[a,b,c,d]

[a]*Theoretical and Scientific Data Science group,*
*Scuola Internazionale Superiore di Studi Avanzati (SISSA),*
*via Bonomea 265, 34136 Trieste Italy*

[b]*Astrophysics Group, Physics Department, Blackett Lab, Imperial College London,*
*Prince Consort Road, London SW7 2AZ, U.K.*

[c]*INFN — National Institute for Nuclear Physics, Via Valerio 2, 34127 Trieste, Italy*

[d]*Italian Research Center on High-Performance Computing, Big Data and Quantum Computing,*
*Casalecchio di Reno, Italy*

*E-mail:* kkarchev@sissa.it, rtrotta@sissa.it

ABSTRACT: Accounting for selection effects in supernova type Ia (SN Ia) cosmology is crucial for unbiased cosmological parameter inference — even more so for the next generation of large, mostly photometric-only surveys. The conventional "bias correction" procedure has a built-in systematic bias towards the fiducial model used to derive it and fails to account for the additional Eddington bias that arises in the presence of significant redshift uncertainty. On the other hand, likelihood-based analyses within a Bayesian hierarchical model, e.g. using MCMC, scale poorly with the data set size and require explicit assumptions for the selection function that may be inaccurate or contrived.

To address these limitations, we introduce STAR NRE, a simulation-based approach that makes use of a conditioned deep set neural network and combines efficient high-dimensional global inference with subsampling-based truncation in order to scale to very large survey sizes while training on sets with varying cardinality. Applying it to a simplified SN Ia model consisting of standardised brightnesses and redshifts with Gaussian uncertainties and a selection procedure based on the expected LSST sensitivity, we demonstrate precise and unbiased inference of cosmological parameters and the redshift evolution of the volumetric SN Ia rate from $\approx 100\,000$ mock SNæ Ia. Our inference procedure can incorporate arbitrarily complex selection criteria, including transient classification, in the forward simulator and be applied to complex data like light curves. We outline these and other steps aimed at integrating STAR NRE into an end-to-end simulation-based pipeline for the analysis of future photometric-only SN Ia data.

ARXIV EPRINT: 2409.03837

https://doi.org/10.1088/1475-7516/2025/07/031

## Contents

## 1 Introduction

Type Ia supernovæ are extremely bright stellar explosions that can be used to study the expansion history of the Universe. In SN Ia cosmology, information about the cosmological model and its parameters is extracted from the relation between apparent brightness, a proxy for distance, and redshift of the supernovæ. However, SNæ Ia are not perfect standard candles and exhibit an intrinsic scatter of absolute magnitudes on the order of 0.1 mag even after standardisation through empirical correlations [1–3]. Therefore, SNæ Ia are subject to Malmquist bias [4, 5], whereby, at a given redshift, less bright members of the population remain undetected in noisy telescope observations (or their properties cannot be reliably determined). This effect increases with redshift, i.e. luminosity distance to the supernova (for a given cosmology), until no dimmer / more distant objects are detected past a certain redshift. Malmquist bias thus results in a difference between the distribution of brightnesses of detected objects with respect to those of the underlying population. Crucially, the size of the bias changes with redshift in a way that can be mistaken for an alternative cosmological model: i.e. it biases the cosmological inference if unaccounted for.

Malmquist bias is but one of numerous *selection effects* impacting SN Ia cosmology, which also include the preference for detecting bluer SNæ Ia or those that remain bright for longer (and thus have a larger chance of being observed). Since these properties are correlated with the supernova's intrinsic brightness, they too have an effect on the inferred cosmology. Furthermore, the data releases of modern observational campaigns and subsequent cosmological analyses employ increasingly sophisticated criteria for detecting, selecting, and classifying SNæ Ia, taking into account the availability and quality of spectroscopic follow-up of the supernova and/or its host galaxy, the fidelity of laborious light-curve model fits [e.g. SALT; 6–9], and the outputs of black-box neural network classifiers [e.g. 10]. This makes an exact treatment of selection effects from first principles practically impossible, necessitating the development of approximate schemes for mitigating the biases induced by selection, which fall into two major categories.

On the one hand, methods for *bias correction* [11] introduce an offset to the observed SN Ia magnitudes that negates the selection bias *on average*,[1] given a fiducial model that fixes e.g. the correlation coefficients between brightness and stretch/colour, the SN Ia rate with redshift, and, most importantly, the cosmological model. The size of this magnitude correction as a function of redshift[2] is fixed and determined from a large preliminary simulation of the SN Ia population to which the full selection procedure is applied.

However, the correction does depend on the chosen parameters and, more importantly, is completely degenerate with the sought-after signature of the cosmological model, namely the variation of brightness with redshift. While the adoption of constraints from external data, e.g. the cosmic microwave background (CMB), may break the degeneracy, in general, bias corrections with incorrect fiducial parameters lead to incorrect cosmological inference (see e.g. figure 4 in [11] and appendix C.3 below). The consistency of analyses with bias correction was recently investigated by [13] in the *frequentist* context with a Neyman construction (similar in spirit to the one we introduced in SICRET, figure 8). However, they do not consider the possibility of significant redshift uncertainties (whose importance we demonstrate throughout this work). Their suggested amendment to the bias correction would fail if the assumed fiducial cosmology is inconsistent with the true one. The required degree of closeness between the sought-after true cosmology and the fiducial cosmology used for bias correction simulations depends on the statistical constraining power of the data and thus, ultimately, on the number of detected SNæ Ia, which is set to increase drastically in coming years. The importance of a principled, transparent and easily reproducible framework is demonstrated by the recent arguments in the literature about an apparent inconsistency of ∼0.04 mag between the common SNæ Ia in the Dark Energy Survey (DES) 5-year sample and in the Pantheon+ compilation, as pointed out by [15]. A rebuttal [16] reproduces the offset but argues that it is due to improved intrinsic scatter modelling and different selection functions between the two samples, which result in different bias corrections.

---
[1]Alongside correcting the mean brightness-redshift trend, the assumed uncertainty around it can also be modified to take into account the reduction in the variety of *observed* SNæ Ia with respect to the total population: see [11, subsection 5.3].

[2]Since [11], the framework for bias corrections has been altered and augmented, notably by [12], and currently encompasses several variants with different "independent" and "response" (i.e. corrected) variables beyond redshift and magnitude: namely, the SN stretch and colour and properties of the host galaxy: see figure 1 in [12].

Usually, the magnitude offset for each SN needs to be calculated for a single redshift value, typically the maximum likelihood estimate (MLE), thus ignoring any associated uncertainty. However, only a small fraction of the detected transients in future surveys will have precise spectroscopic redshift estimates (with uncertainty $\approx 10^{-5}$), relying instead solely on photometry (of the transient itself and/or of the host), which is known to produce highly uncertain and non-Gaussian estimates [17, see also 18, figure 1]. As we show in appendix C, significant redshift uncertainties, in combination with a varying rate of SN Ia explosions with redshift, lead to an Eddington bias [19] in the magnitudes of selected SNæ Ia even in the absence of brightness-related selection/detection criteria, which traditional bias corrections are not designed to handle.[3] Even though the systematic bias is small in absolute terms, it will become dominant when compared to the vanishing statistical uncertainties of future large SN surveys.

While bias correction is an *ad hoc* procedure, more principled — Bayesian — frameworks for SN Ia cosmology have also been developed to properly account for uncertainties on multiple levels: population, individual objects, and instrumental noise in a so-called Bayesian hierarchical model (BHM). Initially introduced to SN Ia cosmology by [21], this approach was extended by [22–26] to include a variety of physical and observational effects like correlations with the host galaxy, non-Ia contamination, and probabilistic sample selection.

Inference within a BHM is traditionally performed through Markov chain Monte Carlo (MCMC) sampling of the joint posterior, which requires either explicit evaluation of the full hierarchical likelihood or analytical marginalization over latent variables, which is only possible under simplifying assumptions of Gaussianity of the distributions and linearity. The full likelihood requires specifying the selection probabilities for each SN as a function of all model parameters: object-specific (stretch, colour, redshift, etc.) and global (cosmology, standardisation coefficients, rates, calibration, etc.), and evaluating them at every step in the chain. But since the selection function is unavailable from first principle, current BHM implementations [24–27] resort to fast analytic approximations and/or potentially inaccurate simplifying assumptions for its analytic marginalisation over unobserved objects.

Furthermore, the scalability of likelihood-based analyses is hindered by the necessity (in general) to sample/infer the full parameter space, which grows with the data set size, as each new SN introduces a set of latent parameters which are tractable only for very specific — typically Gaussian and linear — models. State-of-the-art Bayesian light-curve models like BayeSN [28–30] feature tens of parameters *per SN*, and while some of the computational burden in sampling them can be alleviated through the use of modern Bayesian inference techniques [31, 32] or direct numerical marginalisation [e.g. 33], designing an appropriate high-dimensional *approximation* to the selection probability and tuning/training it remains a significant challenge. On the other hand, scientific conclusions are usually drawn from the marginal posteriors for a handful of global parameters — e.g. the subset related to cosmology or to the connection between SNæ and their hosts.

---

[3]Ref. [20] suggested to *correct* for this additional bias as part of the selection effects treatment. However, this requires knowledge of the true cosmological model to correct towards; in other words, it leads to imprinting the fiducial parameters onto the data even more firmly than usual.

To overcome these challenges, particularly in view of future large-scale surveys like the Legacy Survey of Space and Time (LSST) from the Rubin Observatory and the Roman Space Telescope's Wide-Field Infrared Survey, which are forecast to observe hundreds of thousands of transients, we propose simulation-based inference (SBI): a framework that abstracts the BHM into a stochastic forward simulator used to generate realistic mock data while randomly sampling all model parameters from their hierarchical priors. During inference, the simulator is treated as a black box, which has two important consequences. First, SBI can easily incorporate *any* probabilistic effects (e.g. selection, contamination) as long as they can be represented through forward sampling — rather than the explicit evaluation of a likelihood/probability; hence, SBI is uniquely poised to solve realistic selection effects *exactly*. Secondly, SBI naturally produces *marginal* results for the parameters of interest, implicitly integrating out all other stochastic variables in the simulator; typically, this requires much fewer simulations than likelihood evaluations [see e.g. 34, figure 8] and does not explicitly scale with the size of the latent space (i.e. the data size) beyond the linear need to simulate and process all objects.

The recent rise in popularity of SBI in the physical sciences[4] is largely attributable to the adoption of neural networks (NNs) as either likelihood, posterior, or likelihood-to-evidence ratio estimators in the different flavours of neural SBI: NLE, NPE, NRE; see [35, 36] for overviews and an extended comparison. Previous applications in the field of SN Ia analysis include non-neural SBI using approximate Bayesian computation (ABC) [37–39], NPE for cosmological inference from summary statistics [40–44] and for inferring the properties of individual SNæ Ia from interpolated light curves [45, 46], and hierarchical inference (i.e. of both object-specific and population-level parameters simultaneously) using NRE from large fixed-size collections of summary statistics [14, hereafter SICRET] and from raw light curves [47, hereafter SIDE-real]. Transcending inference, [48] adapted marginal NRE for use in Bayesian *model comparison* applied to an array of hierarchical SN Ia population models.

All previously mentioned works found the capabilities of simple fully-connected NNs (so-called multi-layer perceptrons (MLPs)) sufficient. Meanwhile, a large variety of NN architectures are being developed to extract and exploit the information contained in variously structured real-world and industrial data: e.g. images, point clouds, graphs, word sequences (i.e. natural language), and mixtures thereof (i.e. multi-modal data). Particularly relevant to SNæ — and other transients — are recurrent and attention-based (transformer) architectures, which have been applied to classification [10, 49] and deriving informative fixed-size representations [50–52] from light-curve data, i.e. varying-length *ordered sequences* of observations of *a single* transient.[5]

---

[4]Compilations of references can be found on https://simulation-based-inference.org/ and https://github.com/smsharma/awesome-neural-sbi.

[5]A notable alternative approach to handling the cadence variety of SN light curves is to interpolate them onto a fixed grid in time and wavelength: see e.g. [46, 53].

On the other hand, survey catalogues are inherently unordered data *sets*, whose sizes are still *a priori* unknown: i.e. the number of mock SNæ detected is not fixed in the simulator (see subsection 2.2 for further discussion) and may carry useful information for the parameters of interest (see subsection 3.1). Previous applications of SBI [54–57] have considered only modestly sized sets (up to a few hundreds of objects) and regarded their cardinality as uninformative, simply *amortising* the learning over different-sized data via one of two set-based NN architectures: deep sets [58] and the set transformer [59].[6] While the latter is more expressive and more tailored towards capturing correlations between object pairs (and triples, etc. as the depth is increased), the underlying attention mechanism has quadratic scaling in both memory and compute,[7] and even though this can be somewhat mitigated, the simpler deep set architecture has been shown to better aggregate global information from large sets [57, figure 2]. Combined with a favourable linear scaling of memory and computational requirements, deep sets are our choice for SN Ia cosmology from future-sized data containing on the order of $10^5$ SNæ (see subsection 3.1).

Lastly, we survey the few existing efforts in tackling selection effects with SBI. Ref. [63] noted a similarity between detection/selection and truncation (a method that boosts the performance and simulation-efficiency of neural SBI), although their problem was not catalogue based; instead, selection/detection of point sources in input gamma-ray images was *introduced* for the benefit of truncation and solved using four separate ratio estimators. In the field of standard *siren* inference — of the Hubble constant from binary mergers detected through gravitational waves — ref. [64] used rejection sampling in the simulator to produce mock selected catalogues of a size fixed to that of the real data set, which they then compressed into a few automatically optimised *global* summary statistics with which to perform (global) NLE. Ref. [65] followed a similar approach but used an intricate hand-derived summarisation procedure and NRE. Finally, [66] recently demonstrated an application of a neural density estimator as a very flexible emulator for the selection-affected distribution of summary statistics for an *individual object* in a simplified BHM for SN Ia cosmology. This approach, however, requires a high level of precision to allow for combining the NN results over a large collection of SNæ and disregards the information contained in *unobserved* objects (see subsection 2.1).

In contrast to previous works, our method directly delivers marginal results, imposing no restrictions on the simulator's internal structure when it comes to the latent parameters of individual objects or the output size. It can thus handle an arbitrarily complicated selection

---

[6]Here we must note the alternative approach of applying traditional SBI to individual objects and combining the results with some trivial aggregation like summation of log-probabilities over a data set with inherently unknown size [e.g. 60–62]. Crucially, this strategy is only applicable when the observations are independent and identically distributed (i.i.d.), conditionally on the inferred parameters, which is often not the case with marginal inference and may thus force one to learn a much higher-dimensional likelihood than needed. Even though we do learn the full global parameter space for the sake of truncation, our method does not require conditional independence because of the flexible deep set aggregation.

Moreover, combining individual likelihoods is prone to accumulation of the inevitable approximation errors from approximate per-object inference: e.g. [60, figure 8] observed a bias when combining inference from 1000 objects, while [62] were able to scale to (*exactly*) $10^4$ only by training an additional fixed-input-size network to aggregate the likelihood estimates, correcting the inaccuracy.

[7]Indeed, only recently have the context windows of large language models extended beyond 100 000 tokens.

procedure. Furthermore, it extracts the information contained in the data set cardinality, at the expense of learning a more complicated function from a varying-size set input. It thus represents the definitive solution for handling selection effects in simulation-based inference from object catalogues.

This paper is organised as follows. In section 2, we introduce the problem in generality, the caveats of restricted SBI approaches (illustrated in appendix A), and our strategy for overcoming them, optimising training, and boosting inference fidelity dubbed STAR NRE (we provide implementation details in appendix B). Then, in section 3, we develop a simplified model for SN Ia cosmology that captures the essence of relevant selection effects: Malmquist and Eddington bias, elucidating the latter and demonstrating the failure of traditional bias correction in appendix C. We present results from STAR NRE using mock data in section 4 before concluding and discussing outlook for future applications in section 5.

## 2 Inference framework

In this work, we address a problem in hierarchical Bayesian inference common across many fields: deriving posteriors for the parameters of a *population* of objects from measurements of a *non-random subset* of them, determined by a given *selection/detection* procedure. In this section, we introduce the problem and our solution in generality, whereas in section 3 we detail a simplified model for SN Ia cosmology that we use for demonstration.

### 2.1 Probabilistic description of sample selection

We assume a complete population (before selection) consisting of an unknown number $N_{\text{tot}}$ of objects, out of which $N_{\text{obs}}$ (known) have been measured to produce the data set $\left\{\mathbf{d}^i\right\}$, with $i \in \{1, \ldots, N_{\text{obs}}\}$. We assign to each of the $N_{\text{tot}}$ objects a random variable $\mathcal{S}$ that indicates whether the object has been observed ($= \mathcal{S}_{\text{o}}$) or missed ($= \mathcal{S}_{\text{m}}$). For clarity, we will label missed objects with $j \in \{N_{\text{obs}}+1, \ldots, N_{\text{tot}}\}$ and simply write $\mathcal{S}_{\text{o}}^i$ to mean "object $i$ has been observed/selected/detected" ($\mathcal{S}^i = \mathcal{S}_{\text{o}}$) and similarly with $\mathcal{S}_{\text{m}}^j$ for missed/undetected objects.

In this discussion, "missed" are such objects that *could* have been detected but were not, owing to the particular realisation of their latent properties and observational noise. This scenario is relevant to SN Ia cosmology for two reasons. Firstly, data is extracted from telescope images which in principle include *all* SNæ Ia in the field of view; while dim ones go unnoticed, data is recorded, through forced photometry, for all detected events. Secondly, even among detected SNæ Ia, a majority are discarded (forcefully "missed") for cosmological analyses owing to a deemed "poor quality" of the data, e.g. low signal-to-noise ratio (SNR). In this context, the "total population" are all SNæ Ia within the survey sky area[8] that explode during the survey duration, and the "selection procedure" is the full process that determines which objects get included in the final analysis.

Assuming a hierarchical model with parameters collectively labelled $\boldsymbol{\Psi}$, the likelihood is a product of the following terms: the probability densities of recorded data, $\mathrm{p}\!\left(\mathbf{d}^i \,\middle|\, \boldsymbol{\Psi}\right)$;

---

[8]And up to infinite redshift; in practice, assumptions for the population beyond the detection limit do not influence the analysis, so we can set an arbitrary redshift limit, as long as it is well beyond the furthest detectable object for any population parameters allowed by the priors.

the probabilities of detecting each of the $N_{\text{obs}}$ objects, $\text{p}\big(\mathcal{S}_{\text{o}}^i \,\big|\, \mathbf{d}^i, \boldsymbol{\Psi}\big)$; the probabilities of missing the remaining $N_{\text{tot}} - N_{\text{obs}}$ objects, $\text{p}\big(\mathcal{S}_{\text{m}}^j \,\big|\, \boldsymbol{\Psi}\big)$; and the binomial coefficient $\binom{N_{\text{tot}}}{N_{\text{obs}}}$, which accounts for permutation invariance of the labels:

$$\binom{N_{\text{tot}}}{N_{\text{obs}}} \times \left[\prod_{i}^{N_{\text{obs}}} \text{p}\big(\mathcal{S}_{\text{o}}^i \,\big|\, \mathbf{d}^i, \boldsymbol{\Psi}\big) \text{p}\big(\mathbf{d}^i \,\big|\, \boldsymbol{\Psi}\big)\right] \times \left[\prod_{j}^{N_{\text{tot}}-N_{\text{obs}}} \text{p}\big(\mathcal{S}_{\text{m}}^j \,\big|\, \boldsymbol{\Psi}\big)\right].$$

Inserting $\prod_{i}^{N_{\text{obs}}} \text{p}(\mathcal{S}_{\text{o}}^i \,|\, \boldsymbol{\Psi})$ both in the numerator and in the denominator lets us identify two distinct contributions:

$$\underbrace{\prod_{i}^{N_{\text{obs}}} \frac{\text{p}\big(\mathcal{S}_{\text{o}}^i \,\big|\, \mathbf{d}^i, \boldsymbol{\Psi}\big) \text{p}\big(\mathbf{d}^i \,\big|\, \boldsymbol{\Psi}\big)}{\text{p}(\mathcal{S}_{\text{o}}^i \,|\, \boldsymbol{\Psi})}}_{\prod_{i}^{N_{\text{obs}}} \text{p}\big(\mathbf{d}^i \,\big|\, \mathcal{S}_{\text{o}}^i, \boldsymbol{\Psi}\big)} \times \underbrace{\binom{N_{\text{tot}}}{N_{\text{obs}}} \times \prod_{i}^{N_{\text{obs}}} \text{p}\big(\mathcal{S}_{\text{o}}^i \,\big|\, \boldsymbol{\Psi}\big) \times \prod_{j}^{N_{\text{tot}}-N_{\text{obs}}} \text{p}\big(\mathcal{S}_{\text{m}}^j \,\big|\, \boldsymbol{\Psi}\big)}_{\text{p}(N_{\text{obs}} \,|\, N_{\text{tot}}, \boldsymbol{\Psi})},$$

corresponding to the usual data likelihood of the individual objects given that they have been selected (first term) and to the probability of collecting a data set consisting of exactly $N_{\text{obs}}$ objects from a population of $N_{\text{tot}}$ (second term). Naturally, the latter describes the probability of $N_{\text{obs}}$ selection "successes" out of $N_{\text{tot}}$ trials, based on the success/fail probabilities of the individual "events" (which can be different, depending on the observing conditions and/or noise).

Since $N_{\text{tot}}$ is *a priori* unknown, it needs to be marginalised over after being assigned a prior[9] $\text{p}(N_{\text{tot}} \,|\, \boldsymbol{\Psi})$, which leads to:

$$\text{p}(N_{\text{obs}} \,|\, \boldsymbol{\Psi}) = \sum_{N_{\text{tot}}=0}^{\infty} \text{p}(N_{\text{obs}} \,|\, N_{\text{tot}}, \boldsymbol{\Psi}) \times \text{p}(N_{\text{tot}} \,|\, \boldsymbol{\Psi}). \tag{2.1}$$

Two simplifying circumstances are often present in scientific applications: first, the prior model for the population size, $\text{p}(N_{\text{tot}} \,|\, \boldsymbol{\Psi})$, is a Poisson distribution with rate $\langle N_{\text{tot}} \rangle(\boldsymbol{\Psi})$ calculated from the global parameters; and second, the objects are *a priori* indistinguishable, so all selection probabilities $\text{p}(\mathcal{S} \,|\, \boldsymbol{\Psi})$ are equal (in the absence of data). In this case, the marginalisation (which corresponds to a *thinning* of the counts by a factor $\text{p}(\mathcal{S}_{\text{o}} \,|\, \boldsymbol{\Psi})$; see Prekopa's theorem [67]) results in another Poisson distribution with rate scaled by the success probability:

$$\text{p}(N_{\text{obs}} \,|\, \boldsymbol{\Psi}) \to \text{Pois}[N_{\text{obs}} \,|\, \text{p}(\mathcal{S}_{\text{o}} \,|\, \boldsymbol{\Psi}) \times \langle N_{\text{tot}} \rangle(\boldsymbol{\Psi})]. \tag{2.2}$$

We emphasise that, while $\text{p}(N_{\text{tot}} \,|\, \boldsymbol{\Psi})$ may seem like a modelling choice, it represents the assumed size of the SN Ia population, given e.g. the cosmological parameters that enter into $\boldsymbol{\Psi}$. In subsection 3.1, we describe a physically motivated model based on the volumetric rate of SNæ Ia, in contrast to previous expressions from the literature, which have been arbitrarily designed to enable analytic evaluation of eq. (2.1) for a particular selection probability [24, 25].

---
[9]Defined over the non-negative integers (as the sum in eq. (2.1)); the requirement that $N_{\text{tot}}$ must not be smaller than $N_{\text{obs}}$ is encoded in the likelihood: $N_{\text{tot}} < N_{\text{obs}} \implies \text{p}(N_{\text{obs}} \,|\, N_{\text{tot}}, \boldsymbol{\Psi}) = 0$ by the definition of the binomial coefficients.

Thus, we can finally write the likelihood of the model parameters using only the recorded data on the selected objects and their count:

$$\mathrm{p}\Big(\big\{\mathbf{d}^i\big\}, N_{\text{obs}}\,\Big|\,\mathbf{\Psi}\Big) = \left[\prod_i^{N_{\text{obs}}} \frac{\mathrm{p}\big(\mathcal{S}_{\text{o}}^i\,\big|\,\mathbf{d}^i,\mathbf{\Psi}\big)\,\mathrm{p}\big(\mathbf{d}^i\,\big|\,\mathbf{\Psi}\big)}{\mathrm{p}(\mathcal{S}_{\text{o}}^i\,|\,\mathbf{\Psi})}\right] \times \mathrm{p}(N_{\text{obs}}\,|\,\mathbf{\Psi}). \tag{2.3}$$

Of course, the observed data set already carries information about its size, so we will simply write $\mathrm{p}\Big(\mathbf{D} \equiv \big\{\mathbf{d}^i\big\}\,\Big|\,\mathbf{\Psi}\Big)$.

**The difficulties of performing a likelihood-based analysis** with selection effects are many. Firstly, in eq. (2.3) only the term $\mathrm{p}\big(\mathbf{d}^i\,\big|\,\mathbf{\Psi}\big)$ is usually (but not always, if one is interested in marginal analyses) known. It may also be the case that the selection procedure depends only on the observed data, i.e. $\mathrm{p}\big(\mathcal{S}_{\text{o}}^i\,\big|\,\mathbf{d}^i,\mathbf{\Psi}\big) \to \mathrm{p}\big(\mathcal{S}_{\text{o}}^i\,\big|\,\mathbf{d}^i\big)$ and it does not influence inference. However, in general (and e.g. in the SN detection model described in subsection 3.2 in particular), this probability will depend on *latent* object properties that are either measured with noise or not at all (or at least not included in a data release). An example from SN Ia cosmology is that brightness is standardised in the band of the rest-frame maximum ($B$) whereas detectability is based on whichever band the maximum gets redshifted to, which depends on the true redshift $z$, rather than the noisy measurement $\hat{z}$: see subsection 3.2 for more details.

On the other hand, $\mathrm{p}(\mathcal{S}_{\text{o}}^i\,|\,\mathbf{\Psi})$ and $\mathrm{p}(N_{\text{obs}}\,|\,\mathbf{\Psi})$ result from marginalisation, which is often intractable analytically and performed instead via simulations. Crucially, the likelihood formulation requires these expensive estimates to be performed for every proposed value of $\mathbf{\Psi}$ in e.g. a Monte Carlo chain. Instead, such simulations can be more directly and flexibly used in a likelihood-free simulation-based inference framework like the one presented in this work.

## 2.2 Caveats

SBI is a suite of estimation techniques that target different components of a Bayesian inference task. There are multiple approaches to a simulation-based "solution" for the selection-effects model in eq. (2.3). This subsection discusses the caveats that arise when tackling large sets and provides justification for our preferred global varying-cardinality-set-based method.

One obvious idea is to learn the likelihood[10] from a single selected object (i.e. the term $\mathrm{p}\big(\mathbf{d}^i\,\big|\,\mathcal{S}_{\text{o}}^i,\mathbf{\Psi}\big)$), then evaluate and combine it across the full data set. This has numerous advantages: it requires a simpler network, less training, and $\mathcal{O}(N_{\text{obs}})$ times less memory, while allowing inference from future additions to the data (newly observed objects) without re-training. However, the neural estimate needs to be extremely precise since any approximation errors will compound $\mathcal{O}(N_{\text{obs}})$ times. Moreover, learning per-object likelihoods requires joint inference of all global parameters needed to ensure conditional independence between individual objects, which is more difficult than marginal inference.

Inference (and training) should then be performed from entire sets, whose size may not be known *a priori*, especially in the presence of selection. One may still consider conditioning the simulator to output exactly the correct number of selected objects so that all training examples have the same cardinality and a fixed-input-size network can be used. However, the only option to implement this conditioning is usually rejection sampling, whose efficiency reduces as $1/\sqrt{N_{\rm obs}}$ (the spread of plausible data set sizes) and becomes prohibitively low for large populations. The problem can be mitigated by allowing a range in acceptable $N_{\rm obs}$, especially if $p(N_{\rm obs}\,|\,\boldsymbol{\Psi})$ is not extremely informative, and re-sampling to exactly $N_{\rm obs}$ for input into the network — but this carries all the caveats of ABC.

---

[10]Alternatively, one can learn the posterior via e.g. NPE but must then take care to divide out the prior since it is still the individual *likelihoods* that compose: see [60, eq. (10)].

Finally, one might attempt to learn from non-selected simulated data (i.e. as if the data were randomly sampled from the complete population) and subsequently correct the inference using a learnt detection probability, i.e. estimate separately $\prod_i^{N_{\rm obs}} p(\mathbf{d}^i\,|\,\boldsymbol{\Psi})$ (representing inference without accounting for selection effects) and $\prod_i^{N_{\rm obs}} p(\mathcal{S}_{\rm o}^i\,|\,\mathbf{d}^i, \boldsymbol{\Psi})/p(\mathcal{S}_{\rm o}^i\,|\,\boldsymbol{\Psi})$. While easy to simulate training data for, this approach has the disadvantage that the two models (for the data and the selection labels) do not individually represent the observed data set: indeed, $\left\{\mathbf{d}^i\right\}$ is very different from a typical sample of $N_{\rm obs}$ objects from the complete population, and so it will fall outside of the region in data space in which the inference network has been trained, leading to undefined, and often very biased, results. Similarly, the real collection of detection labels $\{\mathcal{S}_{\rm o}^i\}$, representing $N_{\rm obs}$ detected objects, is very different from random $\left\{\mathcal{S}^k\right\}_{k=1}^{N_{\rm obs}}$. And even if training data were made representative through e.g. a truncation procedure, combining the two terms requires extreme precision in the tails, since if selection effects are important, the results from an analysis that does not account for them — i.e. only the $\prod_i^{N_{\rm obs}} p(\mathbf{d}^i\,|\,\boldsymbol{\Psi})$ term — will be strongly biased and need to be corrected: concrete illustrations with a toy model and in our SN Ia case are presented in appendices A.1 and A.2. In practice, this approach is only applicable if the likelihoods are analytically known — or extremely well estimated — in the whole parameter space.

### 2.3 Set-based truncated auto-regressive neural ratio estimation (STAR NRE)

We are interested in inferring only a small subset $\boldsymbol{\Theta}$ of the model parameters, with the rest, $\boldsymbol{\nu}$, integrated out; i.e. with $\boldsymbol{\Psi} \equiv \boldsymbol{\Theta} \cup \boldsymbol{\nu}$, we seek the marginal posterior

$$p(\boldsymbol{\Theta}\,|\,\mathbf{D}) = \int p(\boldsymbol{\Theta}, \boldsymbol{\nu}\,|\,\mathbf{D})\,\mathrm{d}\boldsymbol{\nu} = \int p(\mathbf{D}\,|\,\boldsymbol{\Theta}, \boldsymbol{\nu})\,p(\boldsymbol{\Theta}, \boldsymbol{\nu})\,\mathrm{d}\boldsymbol{\nu}\,. \tag{2.4}$$

**Neural ratio estimation (NRE).** Ref. [68] approaches the problem by approximating the joint-to-marginal (likelihood-to-evidence; posterior-to-prior) ratio

$$r(\boldsymbol{\Theta}\,;\,\mathbf{D}) \equiv \frac{p(\boldsymbol{\Theta}, \mathbf{D})}{p(\boldsymbol{\Theta})\,p(\mathbf{D})} = \frac{p(\mathbf{D}\,|\,\boldsymbol{\Theta})}{p(\mathbf{D})} = \frac{p(\boldsymbol{\Theta}\,|\,\mathbf{D})}{p(\boldsymbol{\Theta})} \tag{2.5}$$

with a neural network (NN) that classifies parameter-data set pairs $(\boldsymbol{\Theta}, \mathbf{D})$ as either *jointly* or *marginally* sampled: i.e. either coming from $p(\boldsymbol{\Theta}, \mathbf{D})$ or from $p(\boldsymbol{\Theta})\,p(\mathbf{D})$. The network is trained on mock data from a simulator that incorporates all modelled effects, including the selection procedure: thus, naturally, example data sets shown during training consist only of selected simulated objects.

**Truncated marginal neural ratio estimation (TMNRE).** Ref. [69], a variant of sequential SBI, improves on both simulator efficiency and the required network complexity and training time by steering the simulator to produce data increasingly more similar to the real observations. Concretely, it works in *stages* to restrict/constrain the prior from which $\boldsymbol{\Theta}$ are drawn based on the neural ratio estimate from the previous stage, evaluated on the target data. At each stage, incompatible regions of parameter space are excluded ("truncated") from the priors, whose density is otherwise unmodified within the remaining space, save for re-normalisation.

When truncating, it is usually beneficial to also constrain the prior of nuisance parameters that are not of particular scientific interest, which allows further targeting the training data. To achieve this, the vanilla truncation scheme considers an array of parameter groups $\{\boldsymbol{\theta}_g\}$ (which usually contain just one parameter each) and estimates several independent ratios, replacing $\boldsymbol{\Theta} \to \boldsymbol{\theta}_g$ in eq. (2.5). The prior for each group is then truncated to a rectangular box (or a simple 1D interval) so that only an insignificant fraction of the estimated posterior mass is excluded.

When the parameter groups are *a posteriori* correlated and the product of marginals is much less constrained than the joint posterior, however, this scheme is inefficient and may fail to reduce the training-data variability significantly and thus lead to little improvement between stages. Unfortunately, high-dimensional ratio estimation — required for joint inference and non-rectangular truncation of many parameters — usually suffers from significant inaccuracies and approximation noise when trained with a limited training set.

**Auto-regressive neural ratio estimation (ARNRE).** Ref. [70] overcomes this curse of dimensionality by approximating the necessary high-dimensional ratio using a series of low-dimensional estimators. The set of all inferred parameters $\boldsymbol{\Theta}$— which may include nuisances for the purposes of truncation — is first partitioned into disjoint groups: $\boldsymbol{\Theta} = \cup_g \boldsymbol{\theta}_g$ with $\boldsymbol{\theta}_{g_1} \cap \boldsymbol{\theta}_{g_2} = \emptyset$, which are then ordered. While the ordering may be arbitrary, [70, see figure 3] found that it does affect the estimation accuracy in realistic scenarios and recommended putting the most difficult to infer parameters last in the auto-regressive chain, so that their ratio estimators receive more information. For each group a *conditional* ratio estimator is defined: $r_g \equiv r(\boldsymbol{\theta}_g\, ;\, \boldsymbol{\Theta}_{:g}, \mathbf{D})$, in which "preceding" parameters $\boldsymbol{\Theta}_{:g}$ are considered *given* when inferring $\boldsymbol{\theta}_g$. Their product gives the *joint* posterior:

$$\prod_g r(\boldsymbol{\theta}_g\, ;\, \boldsymbol{\Theta}_{:g}, \mathbf{D}) \equiv \prod_g \frac{p(\boldsymbol{\theta}_g, \boldsymbol{\Theta}_{:g}, \mathbf{D})}{p(\boldsymbol{\theta}_g)\, p(\boldsymbol{\Theta}_{:g}, \mathbf{D})} = \frac{p(\boldsymbol{\Theta} \,|\, \mathbf{D})}{\prod_g p(\boldsymbol{\theta}_g)}. \tag{2.6}$$

Note that this is not the same posterior-to-prior ratio as in eq. (2.5); instead, it is the ratio of the posterior to the product of *marginal* priors for each of the parameter groups: $p(\boldsymbol{\theta}_g) \equiv \int p(\boldsymbol{\Theta})\, d\boldsymbol{\Theta}_{\neq g}$ (where $\boldsymbol{\Theta}_{\neq g} \equiv \boldsymbol{\Theta} \setminus \boldsymbol{\theta}_g$ are the parameters not contained in a given group). While the initial priors are usually uncorrelated,[11] making the integral trivial, the high-dimensional non-rectangular truncation scheme we use (described fully in appendix B) modifies the *region* of integration in a parameter-dependent way, leading to complicated marginal priors, and in general $\prod_g p(\boldsymbol{\theta}_g) \neq p(\boldsymbol{\Theta})$.

Still, the procedure for inference with ARNRE is very similar to marginal NRE. First, each of the ratio estimators is trained with the usual binary cross-entropy (BCE) loss but

---

[11]Although the contrary is not uncommon: in fact, we will be using a 2-dimensional correlated Gaussian prior for SN Ia rates (subsection 3.1) but will be grouping the two parameters into a single group $\boldsymbol{\theta}_g$.

with $\mathbf{\Theta} \to \boldsymbol{\theta}_g$ and $\mathbf{D} \to \mathbf{\Theta}_{:g}, \mathbf{D}$:

$$\mathbb{E}_{\mathrm{p}(\boldsymbol{\theta}_g, \mathbf{\Theta}_{:g}, \mathbf{D})}\left[-\ln \frac{\hat{\mathrm{r}}_g}{1+\hat{\mathrm{r}}_g}\right] + \mathbb{E}_{\mathrm{p}(\boldsymbol{\theta}_g)\,\mathrm{p}(\mathbf{\Theta}_{:g}, \mathbf{D})}\left[-\ln \frac{1}{1+\hat{\mathrm{r}}_g}\right]. \tag{2.7}$$

Then their product, which approximates eq. (2.6), is multiplied by $\prod_g \mathrm{p}(\boldsymbol{\theta}_g)$, which can be estimated from the training samples via low-dimensional kernel density estimation (KDE) or using a modified NN as described by [70], and used for MCMC sampling, since even for relatively low-dimensional $\mathbf{\Theta}$ (e.g. our 7D example) the alternative of sampling from the marginal priors and using eq. (2.6) as weights is prone to very large sampling noise.

### 2.3.1 Conditioned deep set for NRE

Sets have two properties that distinguish them from other data like lists (of observations), images, etc.: they are unordered collections, and their size is a variable *feature* rather than a given constant. Therefore, the neural network used for set-based SBI should not only accommodate but also utilise these properties.

A breakthrough in this area was Zaheer et al.'s representation theorem [58] stating that any function $f : \{x \in \mathbb{R}^d\} \to \mathbb{R}^n$ that takes a set as input (and is thus invariant to the order of the set's elements) can be represented via: an element-wise transformation $\phi : \mathbb{R}^d \to \mathbb{R}^m$ of the set's elements, a summation of the resulting "features" $\{\phi(x)\}$, and a post-processing function $\rho : \mathbb{R}^m \to \mathbb{R}^n$; that is

$$f(X) = \rho\left(\sum_{x \in X} \phi(x)\right). \tag{2.8}$$

A *deep set* then consists of deep — but conventional, i.e. fixed size/order-input — NN components that approximate/learn $\rho$ and $\phi$.

The simplest implementation of set-based NRE derives a fixed-size data-set summary $\mathscr{S} = \hat{\rho}\left(\sum_{\mathbf{d} \in \mathbf{D}} \hat{\phi}(\mathbf{d})\right)$ and passes it to a simple ratio estimator $\ln \hat{\mathrm{r}}(\mathscr{S}, \mathbf{\Theta})$, as in previous NRE applications. However, we found this architecture extremely inefficient and slow to train since the deep set is forced to learn/encode the *whole* posterior (at all $\mathbf{\Theta}$ values), only to be queried at a single point by the ratio estimator.[12]

Instead, we propose a minimal modification that streamlines learning by *conditioning* the deep-set summaries on the particular value of $\mathbf{\Theta}$, i.e. providing it to the featuriser $]\phi$ (as well as to the post-processor):

$$\ln \hat{\mathrm{r}}(\mathbf{\Theta}\,;\mathbf{D}) = \hat{\rho}\left(\mathbf{\Theta}, \sum_{\mathbf{d} \in \mathbf{D}} \hat{\phi}(\mathbf{\Theta}, \mathbf{d})\right). \tag{2.9}$$

This form is inspired by the exact expression for the log-likelihood in the case of conditionally independent data given $\mathbf{\Theta}$ (cf. eq. (2.3)), in which the featuriser learns the per-object likelihood while the post-processor accounts for information in the set cardinality and for the evidence. Importantly, the conditioned deep set remains applicable in the more general case when the objects are exchangeable but not necessarily conditionally independent: e.g. if $\boldsymbol{\Theta}$ represents only a subset of the population parameters.[13] In reality, it is also free to take any "shortcuts" that allow it to fulfill the ratio-estimation task as efficiently and accurately as possible. Finally, we write the ratio estimator for auto-regressive parameter group $\boldsymbol{\theta}_g$ as

$$\ln \hat{r}_g\left(\boldsymbol{\theta}_g \,;\, \boldsymbol{\Theta}_{:g}, \left\{\mathbf{d}^i\right\}\right) \equiv \hat{\rho}_g(\boldsymbol{\theta}_g, \boldsymbol{\Theta}_{:g}, \boldsymbol{\mathscr{S}}_g, N_{\text{obs}})$$

$$\text{with} \quad \boldsymbol{\mathscr{S}}_g \equiv \frac{1}{N_{\text{obs}}} \sum_{i=1}^{N_{\text{obs}}} \hat{\phi}_g\left(\boldsymbol{\theta}_g, \boldsymbol{\Theta}_{:g}, \mathbf{d}^i\right), \tag{2.10}$$

where in the interest of numerical performance, instead of summing, we average the featurised set elements and append the cardinality to the output.[14] Importantly, $\boldsymbol{\mathscr{S}}_g$ are, owing to the *conditioned* deep set architecture, bespoke summaries for each parameter group $\boldsymbol{\theta}_g$, instead of the usual single $\boldsymbol{\mathscr{S}}$ shared by multiple ratio estimators.

---

[12]Deriving a global summary is indispensable for neural posterior estimation (NPE), which relies on producing, from data, a conditioning context for a neural density estimator, e.g. a normalising flow.

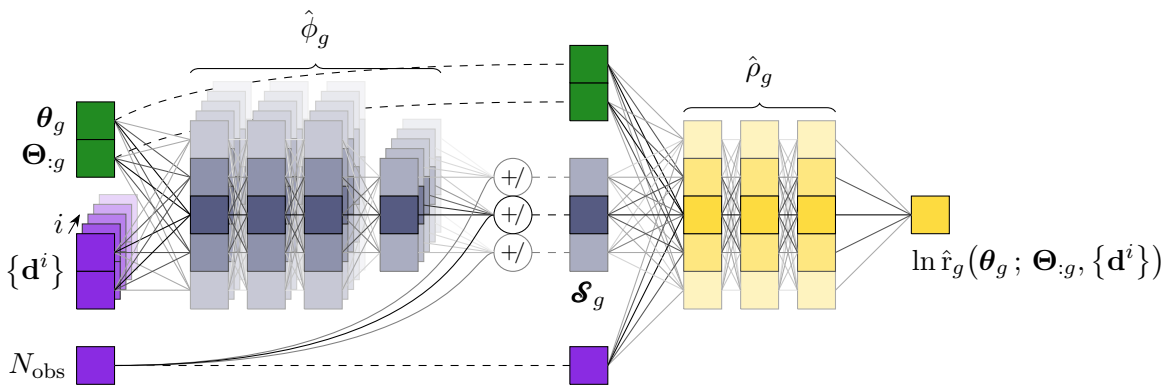

**Figure 1.** Schematic representation of an auto-regressive conditioned deep set-based neural ratio estimator implementing eq. (2.10) with multi-layer perceptrons (MLPs). The architectural details (numbers of layers and their widths) we use in this study are given in section 4.

### 2.3.2 Truncation and inference from (sub)sets

Simulation-based inference with "big data" requires large amounts of time and memory both for simulating training examples and for processing them with a neural network. To alleviate the strain, we propose a truncation scheme particularly suited to data represented as a set and analysed using the deep set architecture. We justify and illustrate it more thoroughly in appendix B; here, we briefly outline the procedure, which relies on the intuition that inference from one small subset of the data is, in general, less constraining than a full analysis, yet unbiased if the subsampling is random — much like intermediate results from early truncation stages.

---

[13]In fact, [58] elucidate this connection between the deep set architecture and hierarchical inference through de Finetti's theorem [71], providing a closed-form expression for the marginal likelihood in the special but widely-applicable case of nuisance parameters with a likelihood from the exponential family and a conjugate prior.

[14]It is good practice to keep inputs to fully connected NN layers close to zero with order-unity scatter, which is the typical range of nonlinearities. However, the output of the summation operator trivially scales with the cardinality of the input set, so we artificially extract this source of variations.

To analyse a given data set of size $N_{\text{obs}}$, we initially train using much *smaller* example/mock sets of size $\approx N_{\text{obs}}/k$ (*nominally*, because each example still has different size), where the "reduction factor" $k$ is arbitrarily set to begin with and iteratively reduced to 1 as we describe below. To simulate such examples, we can either randomly subsample[15] the outputs of a simulator tuned appropriately to represent the full data, or tune it to generate smaller sets if a relevant setting[16] is available. In the latter case, it is important that the setting controls only the total number of objects and not (the distribution of) their properties.

We then train a ratio estimator for the population parameters and evaluate it on $k$ subsets forming a partition of the original data, resulting in $k$ "sub-posteriors" (in fact, posterior-to-prior ratio estimates). Under certain conditions (see appendix B), *multiplying* them *should* give the full posterior(-to-prior ratio) — if the estimates are exact; however, even small inaccuracies might accumulate when combining a large number $k$ of estimates, leading to biased results. Therefore, we adopt a conservative strategy, using an equal-weighted mixture of the sub-posteriors, which we call a "fuzzy" approximate posterior, to truncate the prior: that is, we retain parameters that are consistent with the analysis of *any* of the subsets. We draw constrained samples using MCMC within a volume bounded by multiple 7D ellipsoids, as we describe in detail in appendix B.

We then re-train the ratio estimator on the constrained samples, initialising it with the trained network parameters from the previous stage. If the constraints using the current $k$ do not improve significantly, we switch to a *smaller $k$*, i.e. to generating bigger training sets. In this case, again, we re-use the previous NN parameters. As a result of this iteration between restricting the parameter space from which examples are simulated (i.e. the usual truncation scheme) and increasing the size of example data sets used for training, the two components of the deep set network — the featuriser and post-processor — are quickly and inexpensively pre-trained on small simulations and only fine-tuned on bigger more computationally and memory-intensive ones.

Note that while this procedure serves to determine a suitable truncated prior and initialisation for the network's parameters, the final results are still derived after training on full-size examples and evaluation on the full observed data.

## 3 A simplified model for SN Ia selection effects

This section presents a simplified model for cosmological inference using type Ia supernovæ (SNæ Ia): standardisable candles whose observed properties like light curve shape and colour can be used to infer their *absolute* brightness (up to a certain standardisation uncertainty). Combined with estimates of their redshift and *observed* brightness, this allows one to study the expansion history of the Universe through the distance modulus $\mu(z, \mathcal{C})$, which depends on the cosmological model parametrised by $\mathcal{C}$. We consider $w_0$CDM with parameters the present-day dark matter (DM) and dark energy (DE) densities and the DE equation of state (EOS): $\mathcal{C} \equiv [\Omega_{\text{m}0}, \Omega_{\text{de}0}, w_0]$. In practice, we will make the common further assumption of spatial flatness ($\Omega_{de0} = 1 - \Omega_{m0}$) and fix the Hubble constant, whose value is perfectly degenerate with the mean absolute brightness $M_0$ of SNæ Ia.[17]

---

[15]In fact; partition, taking the big set and assigning each element to a random subset with equal probability, which effectively transforms a small number of large simulations into a larger number of smaller examples.

[16]For astronomical transient surveys, good candidates are the sky area covered and survey duration; in fact, in the setup described in subsection 3.1, these are entirely degenerate.

In order to highlight selection effects and our procedure for handling them, we adopt an intentionally streamlined SN Ia model, considering only the standardised apparent brightnesses $m^s$, traditionally reported in the rest-frame $B$ band at peak, and cosmological redshifts $z^s$ of $N_{\text{obs}}$ observed SNæ Ia, indexed by $s$. They are related to $M_0$ and the distance modulus as:

$$m^s \equiv M_0 + \mu(z^s, \mathcal{C}). \tag{3.1}$$

The data consists of noisy measurements $\hat{m}^s$ and $\hat{z}^s$ of, respectively, $m^s$ and $z^s$. For the magnitudes, we assume constant Gaussian uncertainty that combines intrinsic scatter in the SN population and standardisation and measurement uncertainties:

$$\hat{m}^s \,|\, z^s \sim \mathcal{N}\left(m^s, \sigma_m^2\right). \tag{3.2}$$

For the redshift we assume constant *relative* uncertainty $\sigma_z$, so that

$$\hat{z}^s \sim \mathcal{N}\left(z^s, (1+z^s)^2 \sigma_z^2\right). \tag{3.3}$$

Current cosmological analyses rely on precise spectroscopic redshifts (derived from host spectroscopy even when the SNæ Ia themselves are identified and classified only photometrically) with $\sigma_z \sim 10^{-5}$ and so can assume $\hat{z}^s \approx z^s$. However, properly accounting for the redshift uncertainties will be indispensable when analysing future surveys, as already discussed.

Together, $\mathbf{d}^s \equiv [\hat{z}^s, \hat{m}^s]$ form the data vector for a single observed SN Ia. Their sampling distributions, eqs. (3.2) and (3.3), are controlled by the global/population parameters $M_0$, $\sigma_m$, $\sigma_z$, and the cosmological model. To complete the hierarchical model, which we depict graphically in figure 2, we introduce next the models for the rate of SN Ia occurrence (which determines the population size and redshift distribution) and the selection probability. Finally, we list all involved quantities and their (hierarchical) priors in table 1.

### 3.1 SN Ia rates

The expected redshift distribution[18] of *all* (including undetected) transients of a given type within the surveyed volume, $\Omega T$, where $\Omega$ is the survey sky area (in steradian or deg$^2$), and $T$ is the survey duration (in Earth years), is:

$$\frac{d\langle N_{\text{tot}} \rangle}{dz} = \Omega T \times \frac{R(z)}{1+z} \frac{dV_c(z, \mathcal{C})}{dz}, \tag{3.4}$$

with $dV_c/dz$ the differential comoving volume [72, eq. (28)], which depends on the cosmological model, and the factor $(1 + z)$ accounting for time dilation [73]. For SNæ Ia, we use a comoving volumetric rate $R(z)$ as in the Photometric LSST Astronomical Time-series Classification Challenge (PLAsTiCC) [74, table 2]:

$$R(z) = R_0 \times \begin{cases} (1+z)^\beta, & z \leq z_{\text{break}}; \\ (1+z)^\gamma (1+z_{\text{break}})^{\beta-\gamma}, & z > z_{\text{break}}; \end{cases} \tag{3.5}$$

---
[17] In fact, in our model, $M_0$ represents the combination $M_0 - 5\log_{10}(H_0)$, which is the only way to define a system of units in the absence of distance-ladder rungs calibrated on the nearby Universe.
[18] Technically, the rate function of an inhomogeneous Poisson process.

– 14 –

| parameter | | (hyper) prior | mock value / range | |
|---|---|---|---|---|
| DM density ($z=0$) | $\Omega_{m0}$ | $\mathcal{U}(0,1)$ | 0.3 | |
| DE density ($z=0$) | $\Omega_{de0}$ | $\mathcal{U}(0,1)$ | 0.7 | |
| DE EOS | $w_0$ | $\mathcal{U}(-2,-0.5)$ | $-1$ | |
| rate at $z=0$ | $R_0$ | $\mathcal{N}\left(\begin{bmatrix}2.5\\1.5\end{bmatrix},\begin{bmatrix}0.5^2 & -0.24\\-0.24 & 0.6^2\end{bmatrix}\right)$ | 2.5 | $\times 10^{-5}\,h_{70}^3\,\text{Mpc}^{-3}\,\text{yr}^{-1}$ |
| low-$z$ rate exponent | $\beta$ | | 1.5 | |
| high-$z$ rate exponent | $\gamma$ | fixed | $-0.5$ | |
| rate "break" | $z_{\text{break}}$ | fixed | 1 | |
| true redshift | $z^s$ | $\text{Pois}(d\langle N_{\text{tot}}\rangle/dz)$ | $\in [0;\infty)^{\otimes N_{\text{tot}}}$ | |
| redshift estimate | $\hat{z}^s$ | $\mathcal{N}\left(z^s,(1+z^s)^2\sigma_z^2\right)$ | $\in [0;\infty)^{\otimes N_{\text{tot}}}$ | |
| redshift uncertainty | $\sigma_z$ | $\mathcal{U}(0,0.06)$ | 0.04 | |
| observed magnitude | $\hat{m}^s$ | $\mathcal{N}(M_0+\mu^s,\sigma_m^2)$ | $\in (-\infty;\infty)^{\otimes N_{\text{tot}}}$ | |
| mean abs. mag. | $M_0$ | $\mathcal{U}(-20,-19)$ | $-19.5$ | |
| mag. scatter / noise | $\sigma_m$ | $\mathcal{U}(0,0.2)$ | 0.1 | |
| detection indicator | $\mathcal{S}^s$ | $p(\mathcal{S}^s\,|\,z^s,\hat{m}^s)$ | $\in \{\mathcal{S}_m,\mathcal{S}_o\}^{\otimes N_{\text{tot}}}$ | |

**Table 1.** Quantities in the simplified SN Ia cosmology model. See also figure 2 for a graphical depiction.

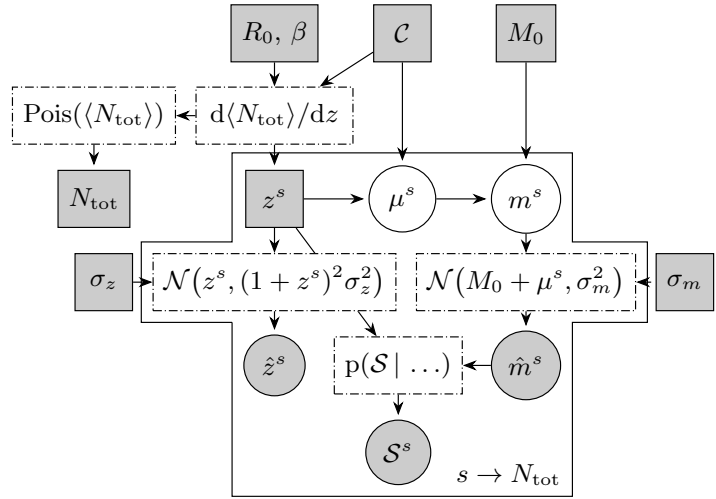

**Figure 2.** Graphical depiction of the simplified SN Ia cosmology model. Shaded squares represent model parameters, sampled according to distributions shown within dashed boxes (some (fixed) parameters and the global (hyper)priors are not shown), while shaded circles are the observed data, and empty circles are deterministic variables.

with slope $\gamma = -0.5$ above $z_{\text{break}} = 1$ [75, after 76, 77], which we keep fixed since LSST is not expected to detect SNæ Ia above $z_{\text{break}}$ for any plausible cosmological model, and so the high-redshift rate model does not influence our analysis. At low redshifts we use the estimates from [73][19] as a 2-dimensional correlated Gaussian prior:

$$R_0 = (2.5 \pm 0.5) \times 10^{-5} \, h_{70}^3 \, \text{Mpc}^{-3} \, \text{yr}^{-1}, ^{\ddagger}$$
$$\beta = 1.5 \pm 0.6,$$
$$\text{with } \text{corr}(R_0, \beta) = -0.8.$$
(3.6)

The rate parameters will be much better constrained with future large samples, as we show in section 4 — assuming the general functional form of eq. (3.5) is correct. However, in upcoming work, we will demonstrate inference with a physically motivated rate of SN Ia occurrence related to the properties of their hosts, which can also inform the progenitor model.

Given particular values for $R_0$, $\beta$, and cosmology, one can sample from the SN Ia population,[20] resulting in $N_{\text{tot}}$ redshift samples, from which observations can be simulated using the sampling distributions eqs. (3.2) and (3.3) described above. Lastly, the collection of selected objects is constructed as described next.

## 3.2 Selection probability

In practice, a SN Ia is detected using difference imaging and selected for inclusion in cosmological analyses based on the quality of its light curve (number of observations in different bands pre/post peak and quality of a fit with e.g. SALT). Since we are building a toy model that only includes the SN's redshift and apparent brightness at peak, our detection criterion will be correspondingly simplified. For the probability of detection & selection (which we call *selection* for short), we adopt a simple criterion based on the expected LSST fiveSigmaDepth for the band in which the SN peaks based on its redshift: at $z = 0$, the peak of SNæ Ia is generally in the blue part of the spectrum; we assume at 4385 Å: the effective wavelength of the $B$ band. The fiveSigmaDepth depends on a variety of factors [78, 79] and in simulations shows significant variations; therefore, for each supernova, we compare $\hat{m}^s$ (which includes both scatter around $M_0$ and observational noise) to a random simulated fiveSigmaDepth,[21] and if $\hat{m}^s$ is lower (brighter), the SN is selected. We thus assume, for the purposes of this toy model, that while the wavelength/band of the peak changes, its magnitude does not; i.e. we omit the so-called *K*-corrections. It is, of course, straightforward to include them in a fuller model in the future.

---

[19]Ref. [73, subsection 6.4.1] give $2.6^{+0.6}_{-0.5}$ but we use 2.5 as [74] and a symmetric uncertainty for simplicity.

[‡]Notice the scaling by $h_{70} \equiv H_0/70 \, \text{km s}^{-1} \, \text{Mpc}^{-1}$, which makes predictions (inference) about total number counts, which depend on the surveyed volume, independent of the particular value of $H_0$ used in simulations (analyses).

[20]To sample from the inhomogeneous Poisson process, we approximate it as piece-wise constant within bins of size $\Delta z = 0.01$ in the interval $z \in [0; 2]$; i.e. we tabulate eq. (3.4) in that range. Within each bin, we then calculate the "integrated" expectation $\left(\frac{d}{dz}\langle N_{\text{tot}}\rangle\right)\Delta z$, sample the number of SNæ Ia from a Poisson distribution, then sample that many redshifts uniformly inside the bin. This results in, approximately, $p(z) \propto \left(\frac{d}{dz}\langle N_{\text{tot}}\rangle\right)$ and $N_{\text{tot}} \sim \text{Pois}\left(\int_0^2 \left(\frac{d}{dz}\langle N_{\text{tot}}\rangle\right) dz\right)$. Note that the upper bound for the complete population does not influence inference results as long as it is much larger than the redshift of any object detectable given any of the *a priori* allowed global parameters.

[21]We use the baseline_v2 run of the LSST operations simulator [OpSim; 80], but the survey details do not affect this simplified model.

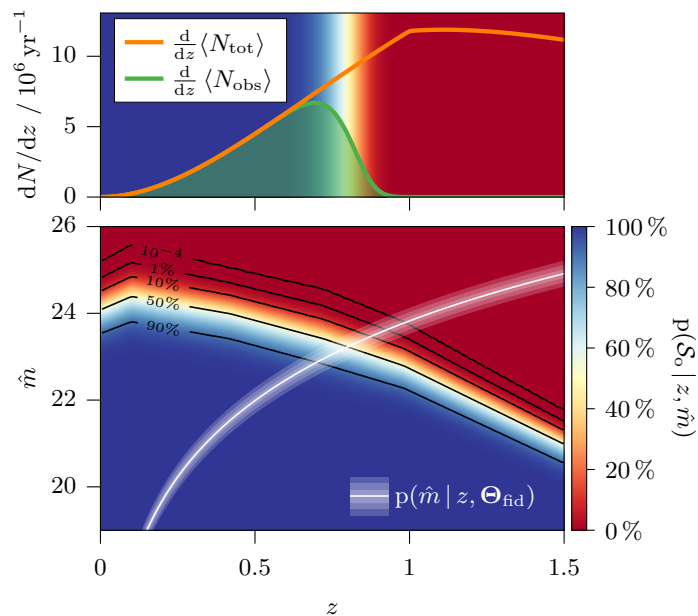

**Figure 3.** *Bottom:* the detection/selection probability of a SN Ia adopted in this work $\mathrm{p}(\mathcal{S}_\mathrm{o}\,|\,z,\hat{m})$, as a function of the SN's true redshift and measured magnitude (colour axis) with some threshold contours as labelled. A white line indicates the relationship between the two variables under the fiducial cosmological model, with up to $3\sigma_m$ uncertainty/scatter around it. *Top:* number (density per unit redshift) of SNæ Ia under the fiducial cosmological and rate models as a function of their true redshift: expected total number in orange and expected detected count in green. The backdrop depicts their ratio $\mathrm{p}(\mathcal{S}_\mathrm{o}\,|\,z,\boldsymbol{\Theta}_\mathrm{fid}) \equiv \int \mathrm{p}(\mathcal{S}_\mathrm{o}\,|\,z,\hat{m})\,\mathrm{p}(\hat{m}\,|\,z,\boldsymbol{\Theta}_\mathrm{fid})\,\mathrm{d}\hat{m}$, i.e. the average along all $\hat{m}$ of the bottom plot weighted by the white shaded region.

Due to the stochasticity of the depth simulator, it defines a complicated selection probability, $\mathrm{p}(\mathcal{S}_\mathrm{o}\,|\,z,\hat{m})$ that depends on $\hat{m}$ and $z$— notice, the *true* redshift rather than the noisy estimate $\hat{z}$. We illustrate it, in comparison with the fiducial distribution of SNæ Ia, in figure 3. Thus, in general, $\mathrm{p}(\mathcal{S}_\mathrm{o}^s\,|\,\hat{z}^s,\hat{m}^s) \neq 1$, even for data on objects that ended up being selected; instead, calculating it requires averaging $\mathrm{p}(\mathcal{S}_\mathrm{o}^s\,|\,z^s,\hat{m}^s)$ over the *posterior* $\mathrm{p}(z\,|\,\{\hat{z}^s,\hat{m}^s\})$, which makes it practically intractable even in this extremely simplified scenario because of the significant redshift uncertainty assumed. However, the selection procedure can be easily realised in a forward simulator: given a collection of (true) redshifts and magnitudes generated as described above, each object is stochastically detected/selected with probability $\mathrm{p}(\mathcal{S}_\mathrm{o}^s\,|\,z^s,\hat{m}^s)$; after all, the simulator always has access to the necessary latent variables.

For real observations, the selection procedure will be far more complex [see e.g. 81, subsection 11.2.1], based on criteria like number of detections in a variety of bands, efficiency of obtaining reliable redshift estimates, correctly classifying transients using machine-learning models, etc., and will be only representable through simulations: the essential impediment to likelihood-based techniques, which our simplified model already exhibits and which we naturally overcome through SBI.

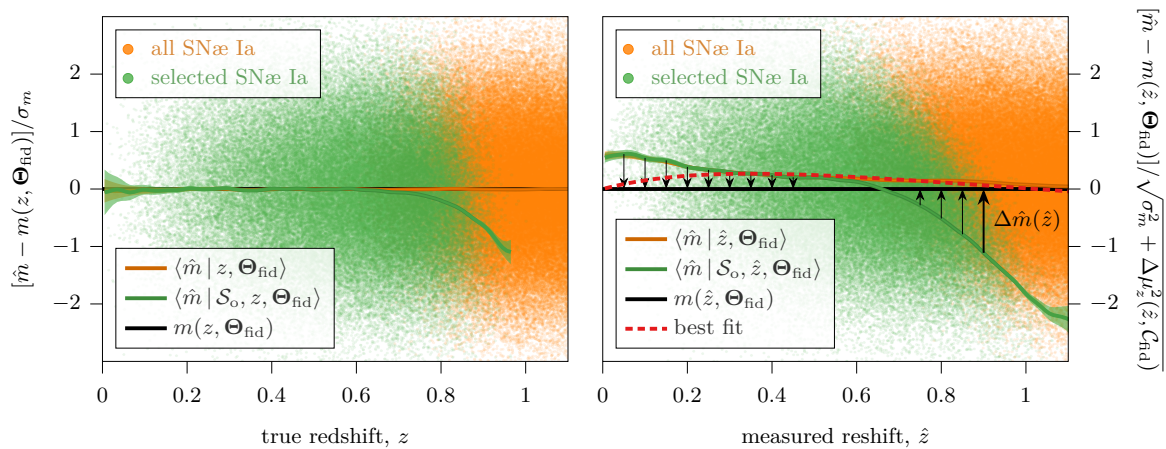

**Figure 4.** The mock data set (selected SNæ Ia and the complete population shown in green and orange, respectively), demonstrating selection bias of the average observed brightness (solid lines) above $z \approx 0.6$. The left panel uses the *true* redshifts (equivalent to spectroscopic estimates: $\sigma_z \approx 0 \implies \hat{z} \approx z$), while the "measured" redshifts used in the right panel are scattered by $0.04 \times (1+z)$, as a simple model of photo-$z$ uncertainties. The ordinate in both cases is the observed magnitude $\hat{m}$ shifted by the fiducial model and normalised by the magnitude scatter, $\sigma_m$, augmented by linearly propagated redshift uncertainty in the right panel (see eq. (C.4)). Note the additional offset in the average brightness of the *complete* sample (orange line) in the presence of redshift scatter, which is due to a varying SN Ia rate with redshift (we demonstrate this Eddington bias-like effect in appendix C). This can lead to biased inference even in the absence of Malmquist bias, as we already showed in SICRET: the best-fit model resulting from a linearised analysis (appendix C.2) of the complete data is plotted as a red dashed line. To prevent this, traditional bias correction (indicated by black arrows) shifts observed data, on average, directly onto the fiducial cosmological model, so that the green and black lines are forced to match under the chosen fiducial model.

### 3.3 Mock data

To demonstrate our inference procedure, we generate a mock data set from the aforementioned model with parameter values (listed in table 1) representing realistic magnitude scatter+noise and redshift uncertainty as expected from a purely photometric measurement. Since LSST is estimated to detect on the order of $10^5$ SNæ Ia/yr[22] [81, figure 11.2], we adjust the survey size[23] so that a similar number (105 287) of mock SNæ Ia are detected under the fiducial rate model and cosmology.

The mock data (apparent magnitudes), for both selected and missed SNæ Ia, are plotted in figure 4 against their true and estimated redshift. In both cases, the selected sample demonstrates the expected bias in favour of brighter objects. However, whereas the complete sample is unbiased when binned according to true redshift $z$, redshift uncertainty ($\sigma_z > 0$) introduces an additional Eddington-like[24] bias [19], whereby SNæ with different true $z$ are measured to have the same $\hat{z}$ (corresponding to a spread in the "posterior" distribution $p(z \,|\, \hat{z})$). This additional shift is present even in the complete population (orange line) and has a detrimental effect on non-hierarchical inference of bias-corrected data (i.e. offset by the $\Delta\hat{m}$ in figure 4), as discussed in appendix C. There we demonstrate that such analyses can only recover cosmological parameters without significant systematic bias if the assumed fiducial model is consistent with the (unknown) true cosmology to within the statistical uncertainty of the analysed data. If this is not the case, the inferred cosmology is systematically biased towards the assumed one. This conclusion is in line with the results obtained by [13], who investigated the shift in cosmological parameters inference induced by an incorrect assumption for the fiducial cosmology used for the bias correction. Our findings show however a much stronger impact, due to our considering redshift uncertainty (ignored by [13]), larger data set size and more discrepant fiducial cosmologies.

---

[22]This is also the size of the mock data we previously analysed in SICRET using TMNRE and a more sophisticated forward model but without selection effects. It is straightforward to include in the present analysis other light-curve summary statistics like SALT's $x_1$ and $c$, noisy measurements $\hat{x}_1$ and $\hat{c}$, and selection criteria based on them, in parallel with $\hat{m}$.

[23]$\Omega T \approx 1600 \deg^2 \cdot$ yr, although the correspondence $\Omega T \leftrightarrow N_{\rm obs}$ from this toy example will not hold for more complicated selection procedures; for comparison, the LSST will cover roughly $20\,000 \deg^2$, a smaller fraction of the SN population will pass the selection criteria employed in practice.

## 4 Results and discussion

For the purposes of set-based truncated auto-regressive (STAR) NRE, we split the global parameters into five groups, ordered by increasing level of "difficulty" to infer:[25] $\boldsymbol{\theta}_g \in [M_0, [\Omega_{m0}, w_0], [R_0, \beta], \sigma_z, \sigma_m]$, for which we train conditional ratio estimators using the loss functional in eq. (2.7). For all featurisers $\hat{\phi}_g$ and post-processors $\hat{\rho}_g$ in the conditioned deep set we use simple MLPs with three layers of, respectively, 128 or 256 neurons and rectified linear unit (ReLU) non-linearities preceded by layer normalisation [82].[26] For added expressivity, parameters passed to $\hat{\rho}_g$ are first embedded in 32 dimensions by a small MLP with two hidden layers of 64 neurons. These choices are inspired by our previous applications in SICRET and SIDE-real and can be automatically optimised by standard pipelines for hyperparameter tuning based e.g. on the validation loss in the last truncation stage.

As explained in subsection 2.3.2, we start with examples from 1/50 of the survey (i.e. $\langle N_{\text{obs}} \,|\, \boldsymbol{\Theta}_{\text{fid}} \rangle \approx 2000$), then truncate the parameter space, preserving only values consistent with a "fuzzy" posterior formed from analysing 50 disjoint subsets of the original mock data, and fine-tune the network on new, targeted examples. Then we increase the survey size tenfold and repeat, *starting with the previously trained network*. Finally, we repeat using the full survey size, this time doing one extra step of truncation to ensure properly converged results. For each of these (7) stages, we generate 64 000 training examples (with 6400 more used for validation), which takes <30 min, and train on 4 (8 for the last stages) NVIDIA A100 graphics processing units (GPUs) using the Adam optimiser [83] until convergence, usually achieved withing 3 h (per stage).

---

[24]Eddington's original argument concerned object counts, but it is easy to extend to the average value of a response variable, as we elaborate in appendix C.

[25]From most to least constrained with respect to the prior range after an initial non-autoregressive marginal NRE run.

[26]The use of layer- instead of batch normalisation is beneficial for two reasons: first, it is slightly faster since it does not need to track running statistics, and second, because it does not depend on the input data, we can reuse network components across training tasks, i.e. truncation and cardinality stages. Still, we perform one-off data-set "whitening" by shifting and re-scaling all NN inputs (parameters and data) by their respective means and standard deviations from the training set.

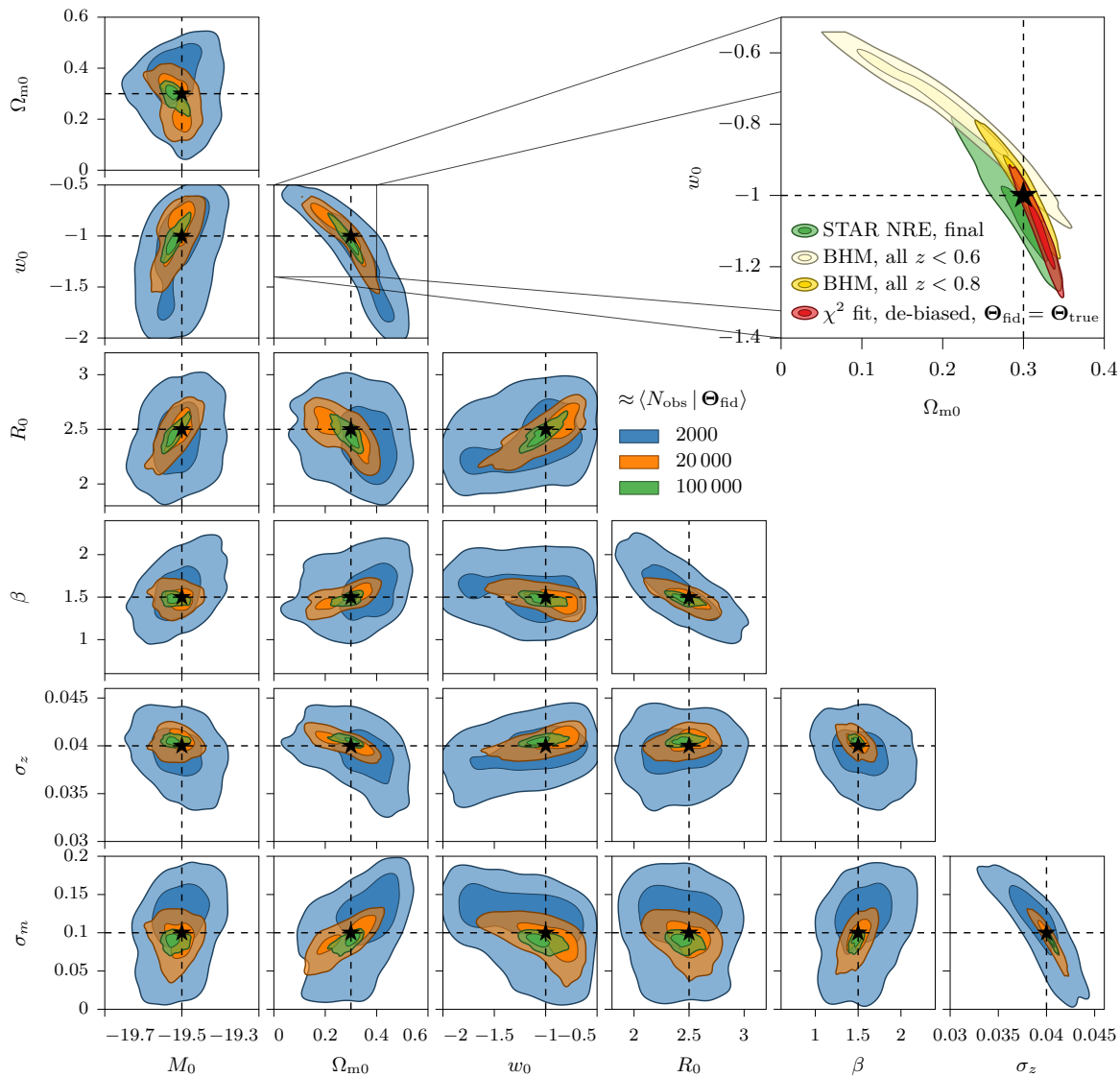

**Figure 5.** Corner plot depicting two-dimensional projections (1- and 2-sigma contours with, respectively, 39 % and 86 % credibility) of the full 7D STAR NRE "fuzzy" approximate posterior (see appendix B). Different colours show the results from successive sub-set-based truncation stages, as described in subsection 2.3.2: the approximate size of analysed sub-sets is indicated by "$\approx \langle N_{\text{obs}} \,|\, \Theta_{\text{fid}} \rangle$" in the legend. A star and dashed line denote the parameters used to generate the mock data. We assume a redshift uncertainty representative of purely photometric estimates. *Inset:* comparison of the final STAR NRE cosmological posterior with likelihood-based analyses of the complete (i.e. volume limited) sample up to $z = 0.6$ and $0.8$ and of the mock selected data de-biased as described in appendix C. Note that inference from the latter is highly dependent on the fiducial parameters used for de-biasing and produces an unbiased posterior only if they match the "true" ones.

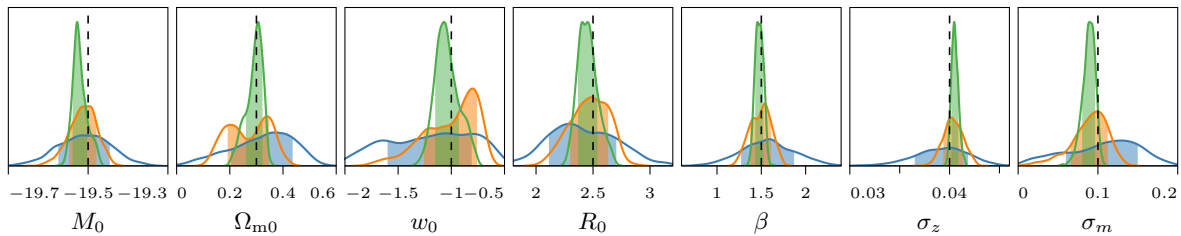

**Figure 6.** One-dimensional marginals of the STAR NRE posteriors in figure 5 for $\langle N_{\text{obs}} \,|\, \boldsymbol{\Theta}_{\text{fid}} \rangle \approx 2000$ (blue), $20\,000$ (orange), $100\,000$ (green). The shaded areas represent the mean $\pm 1$ standard deviation, while dashed vertical lines denote the values used to generate the mock data.

The results from the final stage for each survey size (cardinality stage) are shown in figures 5 and 6. As expected, they are progressively more concentrated around the parameters from which the mock data were generated. To appraise the amount of information extracted by our inference procedure (i.e. the strength of the constraints it produces), in the inset of figure 5, we compare the marginal posterior for the cosmological parameters from STAR NRE to alternative approaches.

On one hand, we perform fully hierarchical Bayesian analysis (with latent redshifts marginalised numerically via integration on a grid) of *complete* data, i.e. that has no magnitude-based selection and all SNæ Ia up to a given cutoff *true* redshift are included.[27] While complete data is naturally more constraining due to the additional presence of high-redshift objects and correspondingly higher statistics, such an analysis would in practice be limited to a low redshift of about $z \approx 0.6$ (in our setup: see figure 3) with the infeasible requirement of selecting the objects based on *true* redshift (so that the analysis is computationally tractable). STAR NRE manages to extract information beyond the completeness limit, effectively correcting the selection bias and delivering constraints comparable to those from the similarly sized complete data (up to $z \approx 0.8$, where the selection probability drops to $50\,\%$).

We also compare our result with a traditional $\chi^2$ fit to selected data that has been "de-biased" as described in appendix C — importantly, with the fiducial model used for calculating bias corrections matching the one from which the mock data were simulated — and analysed with redshift uncertainties propagated linearly to magnitudes, again assuming the fiducial model and disregarding the intrinsic SN Ia rate. Even if the resulting constraints appear more stringent that ours, this is largely due to the optimal choice (made before analysing the data) of a fiducial cosmology close to the true model, which, as we demonstrate in appendix C.3, is indispensable for obtaining systematically unbiased posteriors even for the simplest SN model such as ours. In reality, the selection procedures and bias corrections are far more complicated, making bias-corrected fits prone to systematic bias.

---

[27]This true-redshift cutoff is still a kind of selection criterion, but — as opposed to a cutoff in *measured* redshift — the correction associated with it is a simple re-normalisation of the redshift prior, which we account for by calculating the size of the population below the cutoff and including it in the hierarchical likelihood.

## 5   Conclusions

We have presented a simulation-based analysis framework to handle large data sets affected by selection biases. The cornerstone of our approach is the use of a conditioned deep set neural network as a likelihood-to-evidence estimator, which allows training data to have varying cardinality as a result of stochastic simulation of the underlying population and selection procedure. This enables full freedom in the simulator and straightforward inference contrary to non-set-based methods that condition the simulator on the number of selected objects in the real data (usually achieved through rejection sampling whose acceptance rate vanishes with increasing sample size) or train on unselected mock data and subsequently correct the inference results (very difficult to realise accurately in practice due to a large cancellation between the initial analysis and the selection correction: see appendix A) or analyse individual objects and combine the results post-factum (requiring extreme accuracy when aggregating large samples and an assumption of independence of the set elements conditional on the inferred parameters).

Beyond the network architecture, the method we have developed, dubbed STAR NRE: set-based truncated auto-regressive neural ratio estimation, includes two enhancements over plain NRE intended to facilitate its application. Firstly, we have introduced a procedure that drastically reduces the training time required for inference from large sets by pre-training on random sub-samples, whose size is gradually increased in "sub-sampling stages". This benefits the deep-set featuriser, which learns an object-by-object transformation of the individual data and can thus be trained on small sets. The final result, however, is derived after fine-tuning the network — importantly, the post-processor that aggregates *all* observed objects — on full-size simulations. Thus our methodology imposes no condition of statistical independence on the set elements: i.e. allows even marginal inference.

Moreover, we have employed high-dimensional prior truncation to tailor training examples to the observed data and thus increase simulator efficiency and accuracy of the final posteriors. To this end, we have used an auto-regressive ratio estimator [70] that can learn a relatively large global parameter space (in our case, seven-dimensional) efficiently and precisely. When combining truncation with subset-based inference, we have used (and coined) the notion of a "fuzzy" posterior being an equal-weighted mixture of posteriors derived from disjoint subsets of the data forming a partition. This prevents the accumulation of approximation errors but leads to weaker constraints than properly combining the sub-posteriors (which, however, requires accounting for the dependencies introduced by partitioning the data). Therefore, we use such "fuzzy" constraints only to truncate the prior and draw parameters for generating new training data.

As a demonstration, we have applied STAR NRE to mock data from a simplified SN Ia cosmology setup based on expectations of the LSST. Even though it considers only redshifts and standardised apparent magnitudes (with Gaussian uncertainties on both and in the absence of $K$-corrections) and a selection criterion based purely on brightness at peak — thus setting aside (for the future) numerous complications related to measuring redshift from photometry, detecting and fitting light curves, and light-curve standardisation — our model exhibits the salient features of both Malmquist (brightness preference) and Eddington (abundance preference) biases. While the two have similar influence on the Hubble diagram:

namely, offsetting the average observed magnitudes/distance moduli in redshift bins from the underlying cosmological relation, the latter is rarely explicitly considered in the SN Ia literature, since it arises from the combination of significant noise in the redshift estimates, as expected from future photometric-only surveys, and a non-constant rate of SN Ia occurrence with redshift — and most cosmological analyses to date have relied on spectroscopic samples with negligible redshift estimation uncertainty even if selected only based on photometry.

Furthermore, our model is fully hierarchical and thus easily incorporates (and can be used to extract information about) the volumetric rate of SNæ Ia and its evolution with redshift.[28] This aspect of SN Ia cosmology has remained largely unexplored to date since, while there has been a recent trend towards adopting BHMs, including with selection effects [24–26, 66], most current studies still rely on $\chi^2$-based cosmological fits that — due to linear error propagation — allow no room for modelling a nontrivial redshift distribution and its dependence of cosmological parameters. However, in our study, the evolution of the volumetric rate plays an important role in the simulator, which starts exactly by sampling a redshift population, and is a major driver of the Eddington bias. Finally, joint physics-motivated modelling of the rate of occurrence of SNæ Ia and the stellar populations within their hosts through the so-called delay-time distribution (DTD) has the potential to shed light on the nature of the progenitor systems: a highly debated subject [see e.g. 84, and references therein], but this again requires taking selection effects properly into account.

Another instance in which joint modelling of selection effects in SNæ Ia and their hosts is important was studied by [85]: namely, the complex interplay between the SN Ia stretch and colour parameters used for standardisation (which are affected by a preference for selecting longer lasting and/or bluer SNæ), the corresponding magnitude bias, and the so-called mass step (the difference in absolute magnitude between SNæ Ia hosted in low- and high-mass galaxies). Working in the non-hierarchical framework of bias corrections, they found that an intrinsic correlation between host stellar mass and stretch/colour (indicative of a dependence of the explosion mechanism on the stellar population from which a SN has formed) may manifest instead as a selection-like magnitude offset, thus biasing cosmological inference.

In both examples, the object of scientific interest: intrinsic correlations/trends that hint at the physical model behind SN Ia explosions, was obfuscated by a convoluted inference methodology. In contrast, SBI provides a clear framework for physically meaningful and interpretable analyses by incorporating *a priori* all considered effects into the simulator and performing marginal inference on parameters of interest: e.g. the strength of various host-SN correlations, parameters of the DTD, and, of course, cosmology. Through the use of a set-based methodology like STAR NRE, one can transparently, robustly, and rigorously account for the selection effects affecting such studies.

## 5.1 Outlook

This work represents the final piece of the statistical framework for principled and scalable simulation-based SN Ia inference. What remains is to build a simulator that faithfully represents real transient surveys. We already initiated this with `slicsim`, presented in SIDE-real, that simulates realistic multi-band SN Ia light curves from a hierarchical model including intrinsic chromatic variations, host dust (easily adjustable to include correlations with other host properties), Milky Way (MW) extinction, and a flexible instrument model. Simulated light curves, even if irregularly sampled to match the expected cadence of future observatories, can be input into an inference network after a pre-processing step that uses existing methods like Gaussian process regression onto a fixed grid [e.g. 46, 53], recurrent or attention-based NNs [10, 49–52], or even another layer of deep sets. This circumvents the expensive light-curve fitting stage present in all current analyses and does away with the caveats surrounding the use and interpretation of so-derived summary statistics like stretch, colour, standardised brightness at peak, etc.: e.g. questions of their optimality, assumptions of analytic population distributions, and the difference between latent and measured values. Instead, in the SBI approach, the NN learns to process and summarise the raw data in a way that is optimal for the task at hand, while also allowing inference of the underlying latent parameters, as we demonstrated in SICRET and SIDE-real. Moreover, this end-to-end approach allows inclusion of "low-quality" observations that are currently being rejected based on fit (im)probability and/or anomalous extracted parameter values; the NN can learn to ignore them or else robustly extract the information they carry, however little.

---

[28]Provided a general functional form for it: we use a power law with $1 + z$, but this can easily made more sophisticated.

Simulating full light curves, however, takes much more time than drawing summary parameters from analytic distributions, an issue exacerbated by the need to simulate full data even for objects that end up rejected by the selection procedure.[29] In this respect, *hierarchical* prior truncation, whereby the priors of object-specific model parameters (like individual-SN brightness and stretch) are constrained, can be used to reduce the computational burden.

To complement the SN Ia model, one also needs to include models for various *contaminants*: non-SN Ia transients expected to be detected by non-targeted all-sky surveys. Such are already available due to an ongoing effort in the community: see e.g. SNANA [86] and sncosmo [87]. We intentionally omitted contaminants from this study for the sole reason that classification (which needs to be simulated when creating training data) is usually performed with full light curves, while we used a simplified model of peak brightnesses only. Beyond realistically simulating them according to physically motivated prior rates — which, in parallel to those of SNæ Ia, may have inferrable/free parameters — and subsequently including in the "selection" procedure a "classification" step, contaminants require no modifications to the inference procedure: those that fail to be rejected by the classifier will form a small fraction of the training data sets, and the neural network will learn to better classify and ignore them, or — given that current purpose-made classifiers are already near optimal — learn to derive calibrated posteriors corrected for the bias that contamination causes [see e.g. 88].

Lastly, the simulator needs to consistently model "adjacent" data, the majority of which is photometric observations of SN hosts.[30] In the usual approach, these are used to explicitly derive parameters that affect the SN analysis: redshifts and host stellar mass (which has been found to correlate with SN brightness, stretch, and colour, as discussed). Moreover, host observations can be used to constrain the interstellar dust extinction that the SN is also subjected to. Finally, a small fraction of future detected transients will have spectroscopic follow-up (and some fraction of hosts will, too, e.g. in areas where the LSST overlaps with all-sky spectroscopic surveys like Dark Energy Spectroscopic Instrument (DESI)) and hence, spectroscopic redshifts (and more precisely inferred host properties). These subsets trade off quantity for quality and may prove useful as e.g. low-redshift anchors (similarly to the external surveys included in the DES analysis [89]). However, full utilisation of these data will hinge on accounting properly for the severe selection effects present in them.

---

[29]SBI is not free lunch, and it still needs to represent *all* aspects of the hierarchical model, including the two marginal quantities from eq. (2.3) highlighted in the paragraphs below it, through simulations.

[30]Adjacent to host measurements is the issue of host (mis-)*identification*: a complicated procedure that is known to also cause incorrect inference if left unmodelled [88]. Naturally, this step must also be included in the simulator.

In a SBI setting, inference from adjacent data is *implicit*; instead, all relevant processes are simulated simultaneously, and the network is presented with the totality of (mock) observations and asked to provide marginal (e.g. cosmological) results. In principle, it can thus extract *all* relevant information from the data (provided it is faithfully modelled in the simulator) in the most efficient way (provided an appropriate network structure). A downside of this black-box approach is interpretability, for which efforts towards NN introspection may prove valuable. Finally, to take advantage of heterogeneous data — i.e. a fraction of SNæ with additional spectroscopic constraints, — one can resort to multi-modal architectures [90] that enjoy popularity in AI assistants and have recently been brought to SN science by [52]: they include components designed to process different forms of input (e.g. a light curve, a spectroscopic redshift measurement, host broadband photometry, an image of the host) and then combine the information from any and all available sources via a flexible (usually attention-based) mechanism.

With the above elements falling into place to complement STAR NRE within the broader, flexible framework of simulation-based inference, we believe a robust and principled end-to-end pipeline for cosmological inference from very large photometric data sets will soon be ready to reap the full benefits of upcoming transient surveys.

## Acknowledgments

We are grateful to Christoph Weniger and Anais Möller for fruitful discussions and to the anonymous referee for an in-depth read and useful suggestions. Computations were performed on Cineca's Leonardo cluster, using: `Clipppy`[31] for defining the probabilistic model, the conditioned deep set, and the learning objective; `phytorch`[32] for cosmographic calculations; `netCDF` for storing varying-size data sets; `PyTorch` [91] with `PyTorch Lightning` [92] for training the neural network; `emcee` [93], `SciPy` [94], and `dynesty` [95] in various steps of the truncation procedure.

RT acknowledges co-funding from Next Generation EU, in the context of the National Recovery and Resilience Plan, Investment PE1 — "Project FAIR Future Artificial Intelligence Research". This resource was co-financed by the Next Generation EU [DM 1555 del 11.10.22]. RT is partially supported by the Fondazione ICSC, Spoke 3 "Astrophysics and Cosmos Observations", Piano Nazionale di Ripresa e Resilienza Project ID CN00000013 "Italian Research Center on High-Performance Computing, Big Data and Quantum Computing" funded by MUR Missione 4 Componente 2 Investimento 1.4: Potenziamento strutture di ricerca e creazione di "campioni nazionali di R&S (M4C2-19)" — Next Generation EU (NGEU).

---

[31]https://github.com/kosiokarchev/clipppy.
[32]https://github.com/kosiokarchev/phytorch.

## A On a large cancellation in the likelihood-based treatment of selection effects

In this appendix, we demonstrate the large cancellation in the selection-affected likelihood,

$$\prod_i^{N_{\rm obs}} {\rm p}\big(\mathbf{d}^i\,\big|\,\mathcal{S}_{\rm o}^i, \mathbf{\Psi}\big) = \prod_i^{N_{\rm obs}} \frac{{\rm p}\big(\mathcal{S}_{\rm o}^i\,\big|\,\mathbf{d}^i, \mathbf{\Psi}\big)}{{\rm p}(\mathcal{S}_{\rm o}^i\,|\,\mathbf{\Psi})} \times \prod_i^{N_{\rm obs}} {\rm p}\big(\mathbf{d}^i\,\big|\,\mathbf{\Psi}\big), \qquad (A.1)$$

that renders impracticable the piecewise SBI approach of learning a model unaffected by selection effects and subsequently correcting it. We illustrate it in two cases: a Gaussian toy model specifically designed to highlight our argument, and an even more simplified model for SN Ia cosmology as in section 3 but with perfect knowledge of redshifts.

Our argument concerns, on one hand, the usual data likelihood: ${\rm p}\big(\big\{\mathbf{d}^i\big\}\,\big|\,\mathbf{\Psi}\big)$, and on the other, the selection correction: ${\rm p}\big(\{\mathcal{S}_{\rm o}^i\}\,\big|\,\big\{\mathbf{d}^i\big\}, \mathbf{\Psi}\big)\big/{\rm p}(\{\mathcal{S}_{\rm o}^i\}\,|\,\mathbf{\Psi})$. As the data set grows and the distribution of observed objects becomes markedly different from that of the complete population, constraints on $\mathbf{\Psi}$ derived from just the former will be increasingly biased. Meanwhile, the strength of the correction (compounded $N_{\rm obs}$ times) will also increase to counteract the bias and shift the maximum likelihood region towards the true parameters. Moreover, the range of log-values that each of these terms exhibits across the parameter space naturally scales linearly with the number of observed objects, whereas the region of high combined likelihood only shrinks with its square root, which means that the problem is only exacerbated for large data sets.

Consequently, if the two terms are to be approximated separately, e.g. by training an inference network with mock data representing the complete population and then correcting its result with an approximation of the selection probability, the two estimates must be extremely precise in the tails of both, where the high-combined-likelihood region lies, which is extremely hard to achieve with SBI which learns from examples and hence is most accurate in high-density regions.

### A.1 Gaussian toy model

The setup in this appendix is simple. Data is drawn from a normal distribution $d^i \sim \mathcal{N}(\mu, \sigma^2 + \varepsilon^2)$ and subjected to the selection criterion $d^i > 0$. We wish to infer the parameters $\mathbf{\Psi} \equiv [\mu, \sigma]$ after setting priors ${\rm p}(\mu) = \mathcal{U}(-1, 1)$ and ${\rm p}(\sigma) = \mathcal{U}(0, 1)$ and fixing the "measurement" noise to $\varepsilon = 0.2$.

Given a data set, for which we use a mock realisation with $\mu = 0$, $\sigma = 0.5$, and $N_{\rm obs} = 100$ selected objects, we can easily evaluate all terms in eq. (A.1). The data likelihood is the above-mentioned Gaussian:

$$ {\rm p}\big(d^i\,\big|\,\mathbf{\Psi}\big) = \frac{\exp\left(\frac{-(d^i - \mu)^2}{2(\sigma^2 + \varepsilon^2)}\right)}{\sqrt{2\pi(\sigma^2 + \varepsilon^2)}}. \qquad (A.2)$$

On the other hand, the per-object selection probability ${\rm p}(\mathcal{S}_{\rm o}^i\,|\,d^i, \mathbf{\Psi}) \to {\rm p}(\mathcal{S}_{\rm o}^i\,|\,d^i)$ is independent of the parameters (and unity for any selected objects since the criterion is deterministic).

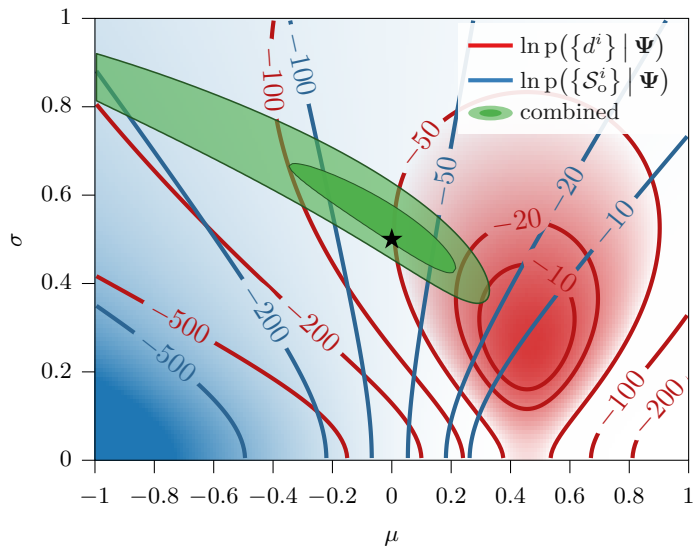

**Figure 7.** Illustration of likelihood cancellation for the Gaussian toy model. The data comprise 100 selected objects ($d^i > 0$) drawn from a Gaussian with $\mu$, $\sigma$ indicated by a star and with added "measurement" noise $\varepsilon = 0.2$. Depicted as blue and red gradients and iso-log-likelihood contours are the two competing terms from eq. (A.1) (normalised to their respective maximum values within the prior range). The setup is engineered so that the two nearly cancel, leading to a combined likelihood $\mathrm{p}(\{d^i\}\,|\,\Psi)/\mathrm{p}(\{\mathcal{S}_o^i\}\,|\,\Psi)$ (illustrated as green filled 1- and 2-$\sigma$ contours) that peaks in the tail of both. Notice the magnitude of the cancellation: across the 2-$\sigma$ region, where the combined likelihood is within $\mathrm{e}^{-2}$ of the maximum, both terms individually change by a factor $\sim \mathrm{e}^{100}$. Therefore, approximating each separately and then combining them could lead to large numerical inaccuracies.

Finally, integrating the data likelihood within $d^i > 0$ gives the marginal probability of selecting an object:

$$\mathrm{p}\left(\mathcal{S}_o^i \,\middle|\, \Psi\right) = \frac{1}{2}\left[1 + \mathrm{erf}\left(\frac{\mu}{\sqrt{2(\sigma^2 + \varepsilon^2)}}\right)\right]. \tag{A.3}$$

The two terms, combined over the 100 objects, are depicted in figure 7. When the two are taken separately, the true values $\mu = 0$, $\sigma = 0.5$ lie in low-likelihood regions (the tails) with respect to the peak values within the prior range. Combining them, however, leads to an almost perfect cancellation leading to the green contours that represent the posterior of an analysis that takes selection effects into account.

Note that in this example we did not include a model for the size of the total population. If we had, we would have been able to put additional constraints on $\mu$ and $\sigma$ through $\mathrm{p}(N_{\mathrm{obs}}\,|\,\mu,\sigma)$, even if the total population size did not explicitly depend on them: e.g. if it were fixed to $N_{\mathrm{tot}} = 200$. In that case, values $\mu \approx 0$, for which a fraction $\approx N_{\mathrm{obs}}/N_{\mathrm{tot}}$ of the population lies within the selection region, would have been preferred by $\mathrm{p}(N_{\mathrm{obs}}\,|\,\mu,\sigma)$.

### A.2 SN Ia cosmology with selection effects and known redshifts

For the case of SN Ia cosmology, we need to make a further simplification to the model from section 3 in order to evaluate and display the exact likelihood and its two constituents: we

– 27 –

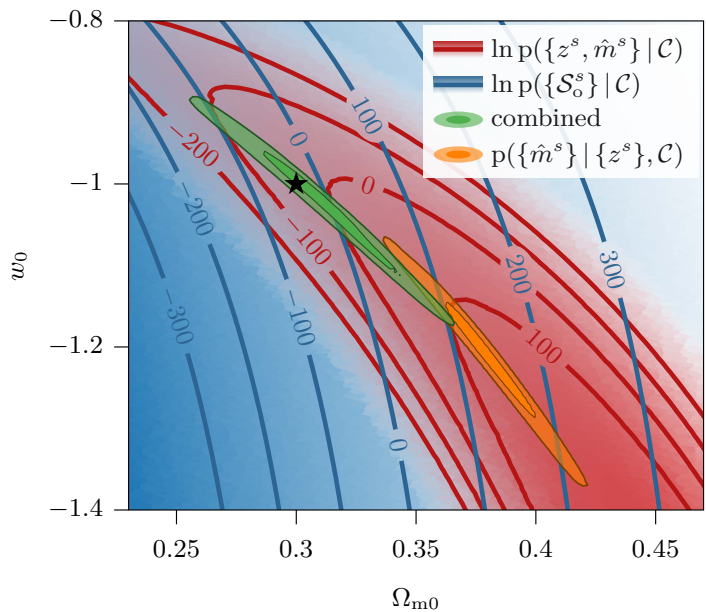

**Figure 8.** Analysis of ≈2000 mock SNæ Ia with the simplified model with negligible redshift uncertainty and selection effects, demonstrating cancellation between the uncorrected data likelihood and the correction for selection effects (red and blue shadings and iso-log-likelihood contours, respectively). The total likelihood, formed by combining these two terms as in eq. (A.1), is depicted as green 1- and 2-sigma filled contours, which lie in the tails of both individual terms. The orange filled contours — systematically biased away from the true parameters (depicted as a black star) — are the result of a naïve analysis which only considers magnitudes (but not the expected distributions of redshifts) and does not account for selection.

assume perfect estimates of the *cosmological* redshifts, which entail negligible measurement uncertainty ($\sigma_z \to 0$), e.g. spectroscopy, and perfect subtraction of peculiar velocities, so that $\mathbf{d}^s \equiv [\hat{z}^s, \hat{m}^s] \to [z^s, \hat{m}^s]$. Therefore, for the data likelihood we have

$$p(z^s, \hat{m}^s \mid \mathbf{\Psi}) = p(\hat{m}^s \mid z^s, \mathcal{C}, M_0, \sigma_m)\, p(z^s \mid R_0, \beta, \mathcal{C})$$
$$\propto \frac{\exp\left[-\frac{\hat{m}^s - (M_0 + \mu(z^s, \mathcal{C}))}{2\sigma_m^2}\right]}{\sqrt{2\pi\sigma_m^2}} \times \frac{R_0 (1+z^s)^\beta}{1+z^s} \frac{dV_c(z, \mathcal{C})}{dz}\bigg|_{z=z^s}. \quad \text{(A.4)}$$

To aid with presentation, we will also fix the nuisance global parameters ($M_0$, $\sigma_m$, $R_0$, $\beta$) to their true values used for simulating the mock data, so that the likelihood only depends on the cosmological parameters of interest: $\mathbf{\Psi} \to \mathcal{C}$.

The per-object selection probability $p(\mathcal{S}_o^s \mid z^s, \hat{m}^s)$ is again independent of the parameters since we assume knowledge of the otherwise-latent $z^s$. However, it still has to be calculated (using the model from subsection 3.2) and multiplied by eq. (A.4) to calculate the marginal probability of selecting a SN Ia:

$$p(\mathcal{S}_o^s \mid \mathcal{C}) = \iint p(\mathcal{S}_o^s \mid z, \hat{m})\, p(z, \hat{m} \mid \mathcal{C})\, dz\, d\hat{m}\,; \quad \text{(A.5)}$$

we perform the integral numerically on a grid of $z$ and $\hat{m}$.

In figure 8, we plot the two terms evaluated for a data set of $\approx 2000$ SNæ Ia (1/50 of the mock data from the main text) as gradients and iso-log-likelihood contours, this time normalised to their respective values at the location of the maximum combined likelihood (near $\Omega_{m0} \approx 0.3$, $w_0 \approx -1$). Here again we see the large range in both the pure-data likelihood and the selection correction and their almost perfect cancellation across the region of high combined likelihood (green contours). For comparison, we also show a naïve analysis that takes into account only the observed magnitudes at fixed redshifts and does not correct for selection effects, i.e. only the term $\prod_s^{N_{\mathrm{obs}}} \mathrm{p}(\hat{m}^s \,|\, z^s, \mathcal{C})$, which is, naturally, biased. Importantly, it also differs from the full data likelihood $\prod_s^{N_{\mathrm{obs}}} \mathrm{p}(z^s, \hat{m}^s \,|\, \mathcal{C})$, highlighting that useful cosmological information is contained also in the distribution of observed redshifts, owing to the volume-related term $\mathrm{d}V_{\mathrm{c}}(z, \mathcal{C})/\mathrm{d}z$ in the observed SN Ia rate. Lastly, we have again ignored the information contained in the number of detected SNæ, which gives much weaker constraints on cosmology than inference from standardised brightnesses with spectroscopic redshifts.

## B  Truncation for joint inference from sets

In this appendix, we explain our implementation of high-dimensional prior truncation, which reduces the diversity of training examples and focuses the learning on the data to be analysed. We also explain our strategy of using inference from random subsets while truncating, in order to optimise simulation and training time and memory requirements by training the inference network on smaller example sets. Nevertheless, for our final result, we want to consider the data set as a whole, and combining inference from subsets is straightforward only under two specific conditions, which we detail below, and prone to the accumulation of approximation errors.[33] Instead, we adopt a conservative approach to subset-based inference applicable regardless of the specifics of the model and use it only for constraining the parameter priors — our final ratio estimator is trained (fine-tuned) on full-size mocks.

### B.1  Combining inference from subsets

Suppose we have a data set $\mathbf{D} \equiv \left\{\mathbf{d}^i\right\}$ with cardinality $\#\mathbf{D} = N_{\mathrm{obs}}$ and a simulator $\mathcal{S} : \mathbf{D} \,|\, \boldsymbol{\Psi}$ by which the data may have realistically been produced (including the cardinality). We can *partition* $\mathbf{D}$ — or any simulated data set — randomly into $k$ subsets $\{\mathfrak{D}_j\}$ by assigning each element to one of the subsets with equal probability. Could one arrive at the combined likelihood from eq. (2.3) via separate $\mathcal{S}$-based analyses of the $\{\mathfrak{D}_j\}$? Yes, but only if one is only interested in global parameters and the model satisfies two conditions.

Firstly, the selected sample size must not depend on the global parameters, and so must the size of the complete population and the selection probability, i.e.

$$\left.\begin{array}{r}\mathrm{p}(N_{\mathrm{tot}} \,|\, \boldsymbol{\Psi}) \to \mathrm{p}(N_{\mathrm{tot}}) \\ \mathrm{p}(\mathcal{S}_{\mathrm{o}} \,|\, \boldsymbol{\Psi}) \to \mathrm{p}(\mathcal{S}_{\mathrm{o}})\end{array}\right\} \implies \mathrm{p}(N_{\mathrm{obs}} \,|\, \boldsymbol{\Psi}) \to \mathrm{p}(N_{\mathrm{obs}}). \tag{B.1}$$

Otherwise, since the cardinalities of subsets forming a partition are *correlated*: $\mathrm{p}(\{\#\mathfrak{D}_j\} \,|\, \boldsymbol{\Psi}) \neq \prod_j^k \mathrm{p}(\#\mathfrak{D}_j \,|\, \boldsymbol{\Psi})$, combining the information contained in them marginally leads to a weaker constraint than using them jointly.

---

[33]For methods that attempt to do this — under different assumptions and in a likelihood-based framework — consult e.g. [96–99].

– 29 –

In our case, the size of the complete set is influenced by the cosmological model through the total cosmic volume being surveyed — and, of course, by the rate parameters — and so is the selection probability through the distance modulus. Therefore, $N_{\text{obs}}$ is informative of the global parameters,[34] and this first condition does not hold, meaning that independent analyses of data subsets cannot extract the full cosmological information of the combined set.

Secondly, even ignoring the information contained in counts one must infer jointly all global parameters that ensure conditional independence of the individual objects so that the likelihoods related to each of them (the term $\text{p}\left(\mathbf{d}^i \,\middle|\, \mathcal{S}_{\text{o}}^i, \boldsymbol{\Psi}\right)$ in eq. (A.1)) compose trivially through multiplication. In other words, while it is common to have i.i.d. observations conditioned on some large set of global/population parameters, the integral from eq. (2.4): $\text{p}(\mathbf{D} \,|\, \boldsymbol{\Theta}) = \int \prod_i \text{p}\left(\mathbf{d}^i \,\middle|\, \mathcal{S}_{\text{o}}^i, \boldsymbol{\nu}, \boldsymbol{\Theta}\right) \text{p}(\boldsymbol{\nu} \,|\, \boldsymbol{\Theta}) \, \text{d}\boldsymbol{\nu}$, may introduce correlations between the objects, and so $\text{p}(\mathbf{D} \,|\, \boldsymbol{\Theta})$ may not factorise. In marginal SBI, this integral is performed implicitly when inferring a small number of parameters of interest. In contrast, ARNRE facilitates inference of (or at least placing constraints on) the whole global parameter space, so that $\boldsymbol{\nu}$ only contains object-specific parameters and conditional independence is ensured.

In any case, combined inference from subsets — even though theoretically possible with ARNRE and ignoring the data set size as a source of information — remains a technical challenge since any small inaccuracies in the (neural ratio) *estimator* employed, possibly insignificant on the level of individual objects/subsets, would compound over the many ($k$) sub-analyses and lead to a biased result.

Therefore, we only use inference from subsets as part of the truncation procedure leading up to an analysis of the complete data set, aiming to be conservative when constraining the parameter space. Thus, instead of the usual product of likelihoods (or likelihood-to-evidence ratio estimates), we use the equal-weighted mixture of posteriors derived from each subset. This results in wide and rough (due to sub-sampling variance), hence "fuzzy", contours, which we use when truncating. The concept is illustrated in figure 9, where samples from the truncated prior (black points) envelop *all* of the sub-posteriors (instead of just their overlap), and while this approach results in truncating less aggressively, it ensures that no parameter regions are incorrectly truncated away.

## B.2 Set-based truncated auto-regressive neural ratio estimation

Finally, we describe the technical implementation of (sub)set-based truncation with "fuzzy" posteriors from ARNRE.

### B.2.1 Simulating subsets

To begin, we need to simulate subsets, for which there are two options. On one hand, we can just simulate complete data sets and partition each into $k$ subsets, resulting in a distribution for the cardinalities of training examples:

$$\text{p}(\#\mathfrak{D} \,|\, \boldsymbol{\Psi}) = \int \text{Binom}(N_{\text{obs}}, 1/k) \, \text{p}(N_{\text{obs}} \,|\, \boldsymbol{\Psi}) \, \text{d}N_{\text{obs}} \,, \tag{B.2}$$

---
[34]In fact, inference purely from the *counts* of detected transients of different kinds, i.e. with different redshift evolution of their volumetric rate, is a promising avenue — if the selection procedure is properly modelled — which we plan to explore in future SBI work.

– 30 –

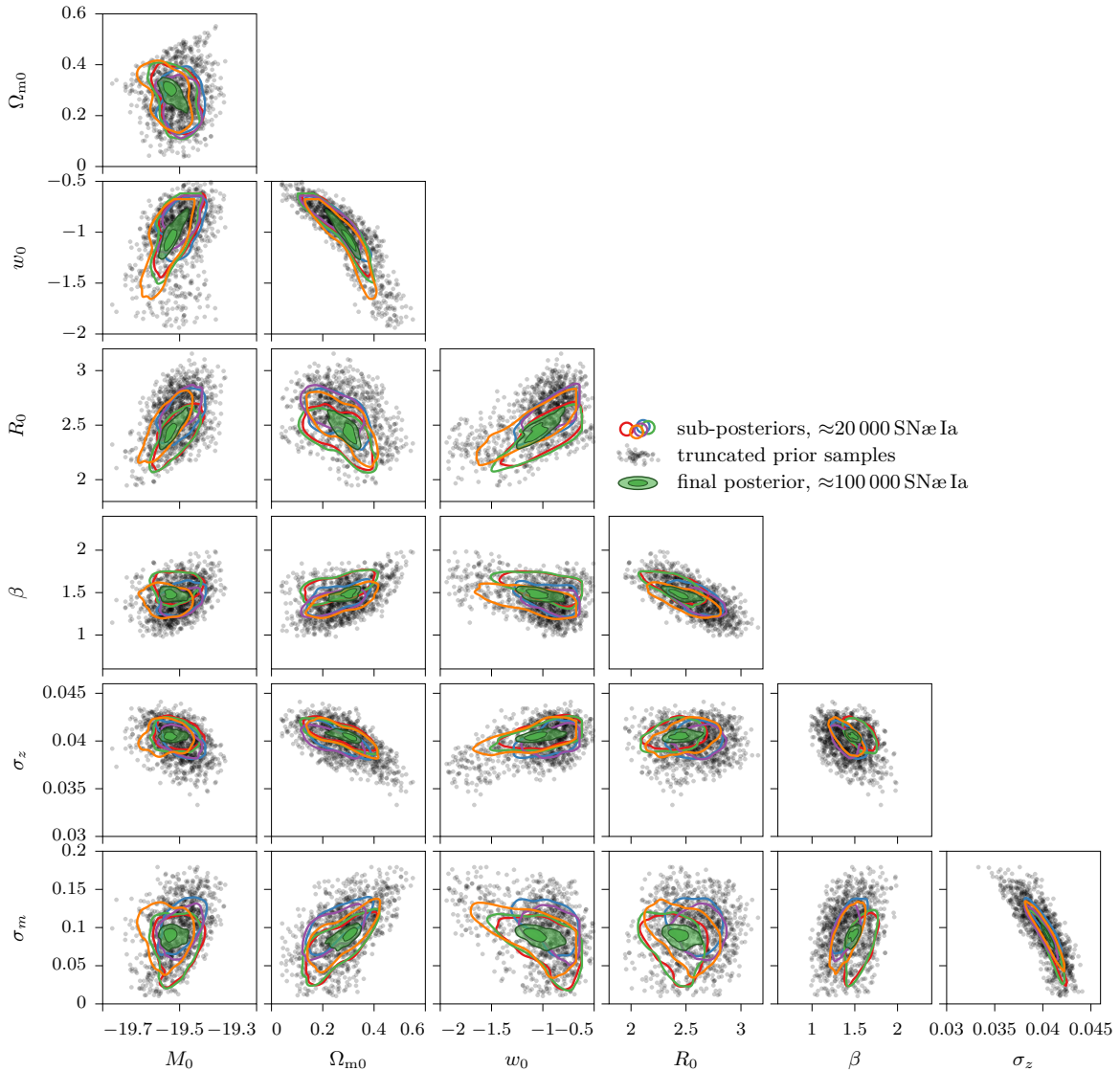

**Figure 9.** An illustration of high-dimensional truncation using a "fuzzy" posterior. Here, the "$\langle N_{\rm obs} \,|\, \Theta_{\rm fid} \rangle \approx 20\,000$" contours from figure 5 are decomposed into five sub-posteriors (only 2-sigma contours shown) resulting from separate analyses of the subsets forming a $k = 5$ partition of the mock data. The black points are drawn from the seven-dimensional prior constrained by a multi-ellipsoid bound encompassing all sub-posteriors. In the subsequent stage, these samples are used to train the final $k = 1$ ARNRE, whose results, after additional truncation, are also reproduced from figure 5.

very similar to the distribution of observed counts from eq. (2.3). As mentioned, if the variable being marginalised — $N_{\rm obs}$ here and $N_{\rm tot}$ in eq. (2.3) — is Poisson distributed, i.e. $\mathrm{p}(N_{\rm obs} \,|\, \Psi) = \mathrm{Pois}(\langle N_{\rm obs} \,|\, \Psi \rangle)$, the result is again a Poisson distribution with an appropriately scaled rate, $\langle N_{\rm obs} \,|\, \Psi \rangle / k$. Since the same holds for the full observed data set as a subset of the total population, subsets $\mathfrak{D}$ can alternatively be directly generated by scaling the total population size in the simulator, i.e. $\langle N_{\rm tot} \rangle (\Psi) \to \langle N_{\rm tot} \rangle (\Psi)/k$. We achieve this by tuning the survey sky coverage and/or duration, which are entirely degenerate and enter into the model only via the product $\Omega T$, so that the simulator directly produces example "sub"-sets, on which we train the ARNRE.

### B.2.2 Sampling from the "fuzzy" posterior

Once the ratio estimator is trained, we partition the "real" data[35] and calculate a collection of sub-ratios, which we sum[36] to form the "fuzzy" posterior-to-prior ratio:

$$\tilde{r}(\boldsymbol{\Theta}\,;\,\{\mathfrak{D}_j\}) \equiv \sum_{j=1}^{k} \frac{\hat{r}(\boldsymbol{\Theta}\,;\,\mathfrak{D}_j)}{k} = \sum_{j=1}^{k} \frac{\prod_g \hat{r}_g(\boldsymbol{\theta}_g\,;\,\boldsymbol{\Theta}_{:g}, \mathfrak{D}_j)}{k}. \tag{B.3}$$

As in eq. (2.6), multiplying this "fuzzy" ratio estimate by $\prod_g p(\boldsymbol{\theta}_g)$, for which we use low-dimensional kernel density estimation (KDE) implemented in `SciPy` [94], gives the full "fuzzy" posterior:

$$\tilde{r}(\boldsymbol{\Theta}\,;\,\{\mathfrak{D}_j\}) \prod_g p(\boldsymbol{\theta}_g) \approx \sum_{j=1}^{k} \frac{p(\boldsymbol{\Theta}\,|\,\mathfrak{D}_j)}{k}. \tag{B.4}$$

We sample from eq. (B.4) using seven-dimensional affine-invariant MCMC implemented in `emcee` [93] for all ARNRE contours shown in this paper: of the "fuzzy" and final posteriors in figure 5 and for the individual sub-posteriors $p(\boldsymbol{\Theta}\,|\,\mathfrak{D}_j)$ depicted in figure 9.

### B.2.3 Constrained prior sampling

Vanilla TMNRE considers low-dimensional marginal inference and usually truncates to a coordinate-aligned box (in one dimension, this is a simple interval). When introducing ARNRE for high-dimensional joint inference, [70] noted the inefficiency of this method for correlated high-dimensional inference and proposed an alternative constrained prior sampler[37] based on slice sampling. In the present work, we adopt a similar approach but make two modifications. Firstly, instead of an iso-(approximate-)likelihood contour, we use the locus of our "fuzzy" posterior samples,[38] approximated using multiple bounding ellipsoids by `dynesty` [95, see also references therein]. And secondly, we sample from the *joint* prior subject to the multi-ellipsoid constraint $\mathcal{B}$ again using `emcee` with probability density proportional to $p(\boldsymbol{\Theta})$ inside $\mathcal{B}$ and 0 outside. We found that affine-invariant MCMC was faster than slice sampling, especially when using a large number of chains that benefit from GPU parallelisation of the ARNRE evaluation,[39] even though we need very long chains to accumulate enough *independent* MCMC samples for the next stage's training and validation sets: to ensure this, we thin the chains by a factor given by the auto-correlation time reported by `emcee`. Once we have the pre-defined number of global-parameter samples, we run the simulator for each of them to produce corresponding data sets of selected supernovæ.

---

[35]In general, each different partition might lead to a different "fuzzy" posterior. However, since we only use it as an intermediate stage in the truncation scheme, we only use one random partition at each $k$-stage.

[36]For numerical stability, and because the network approximates the *log*-ratio, we use the log-sum-exp operation.

[37]https://github.com/undark-lab/torchns.

[38]If the effective MCMC sample size is $n_{\text{eff}}$, this approximates an iso-likelihood contour below/outside which a fraction $1/n_{\text{eff}}$ of the (approximate "fuzzy") posterior mass lies.

[39]While `torchns`'s slice sampler also boasts GPU parallelisation, it requires multiple likelihood evaluations per single sample in order to achieve a lower auto-correlation. Crucially, the number of intermediate steps may be different for each chain, which drastically reduces the efficiency of parallelisation. `emcee`, on the other hand, produces a valid — albeit auto-correlated — sample from every chain with every likelihood evaluation.

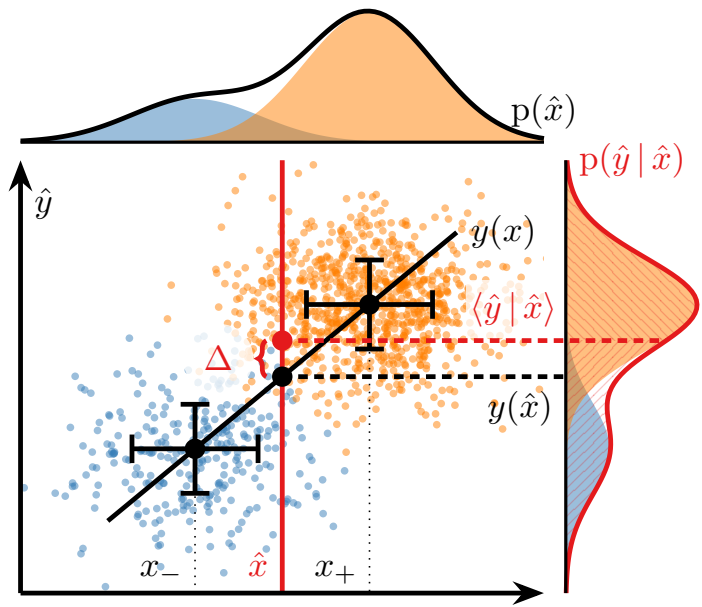

**Figure 10.** Illustration of Eddington bias ($\Delta(\hat{x}) \equiv \langle \hat{y} \,|\, \hat{x} \rangle - y(\hat{x})$) in the response variable ($\hat{y} \equiv y(x) +$ noise) caused by scatter in the independent variable ($\hat{x} \equiv x +$ noise) combined with a trend in its density. For simplicity, here the distribution of the latent $x$ has only two discrete values ($x_-$ and $x_+$) with higher probability at $x_+$. Noisy measurements (here Gaussian with constant variance) are made at a range of $\hat{x}$, and due to the prevalence of objects with true $x_+$, the response variable exhibits Eddington bias towards $y(x_+)$ and away from the underlying $y(\hat{x})$. Such a bias arises in qualitatively the same way for any continuous distribution of $x$ that varies significantly on the scale of measurement uncertainty and independently of the noise in $\hat{y}$, as long as it does not depend on $x$.

## C  Bias correction and redshift uncertainties

### C.1  De-biasing SN brightnesses

The main aim of bias correction in SN Ia cosmology is to counteract Malmquist bias, whereby brighter objects have a higher probability of being detected and selected. As a simple consequence, the average brightness of selected SNæ Ia in bins of redshift[40] is shifted with respect to the total population, as illustrated in the left panel of figure 4. Therefore, a correction

$$\Delta \hat{m}(z) \equiv \langle \hat{m} \,|\, z, \boldsymbol{\Theta}_{\text{fid}} \rangle - \langle \hat{m} \,|\, \mathcal{S}_{\text{o}}, z, \boldsymbol{\Theta}_{\text{fid}} \rangle \qquad \text{(C.1)}$$

meant to undo the shift on average[41] is applied to the observed $\hat{m}$.

---

[40]For more complicated models, the bias can (and does) depend on all other SN-specific parameters like stretch and colour. Since the state-of-the-art correction procedure resorts to histogramming, however, it is difficult to extend beyond a few latent parameters per SN Ia.

[41]In addition, some analyses correct the assumed *scatter* at redshifts where selection effects are important: this results in a *reduction* of the uncertainty in a $\chi^2$ fit (since the SN Ia population is *restricted* to the range

Since calculating the expectation values requires a large number of realistic simulations, $\Delta \hat{m}$ cannot be calculated for every $\boldsymbol{\Theta}$ sampled in a MCMC chain and is instead pre-computed for a fixed *fiducial* choice of global and cosmological parameters $\boldsymbol{\Theta}_{\text{fid}}$. One can then only *hope* that the strength of the selection bias depends weakly on the cosmological model, i.e. that one can already constrain $\boldsymbol{\Theta}$ well enough with data unaffected by selection effects (or with non-SN Ia data altogether, e.g. by including a prior from the CMB).

A further Eddington bias [19] arises — even in the absence of magnitude-based selection — when only noisy redshift estimates are available and the distribution of SNæ is not constant with redshift: we illustrate this effect, which favours *more common* objects, in figure 10 with an idealised setup. Since the SN Ia rate eq. (3.4) increases with redshift and so does the apparent magnitude, more SNæ with a measured $\hat{z}$ have true redshift larger than that, rather than smaller, and so $\langle \hat{m} \,|\, \hat{z}, \boldsymbol{\Theta}_{\text{fid}} \rangle > m(\hat{z}, \boldsymbol{\Theta}_{\text{fid}})$ as seen in the right panel of figure 4. An additional, trivial, source of non-constancy in the redshift distribution is the "lack" of SNæ with $z < 0$, which, as we demonstrated in SICRET, leads to an unavoidable offset at low redshift (seen also in the right panel of figure 4) and incorrect inference even without selection effects and with known redshift distribution.

Eddington bias means that a definition for "photo-$z$ bias correction" equivalent to eq. (C.1):

$$\Delta \hat{m}(\hat{z}) \stackrel{?}{\equiv} \langle \hat{m} \,|\, \hat{z}, \boldsymbol{\Theta}_{\text{fid}} \rangle - \langle \hat{m} \,|\, \mathcal{S}_{\text{o}}, \hat{z}, \boldsymbol{\Theta}_{\text{fid}} \rangle, \tag{C.2}$$

may still not lead to correct cosmological inference, as we show in the right panel of figure 4. Instead, previous instances of cosmological inference from SN Ia samples with photometric redshift estimates [20, 100] have corrected for this Eddington-like shift *together* with the usual Malmquist bias by defining the bias correction not with respect to the average of the complete population but to *the true brightness-redshift relation under the fiducial model*:

$$\Delta \hat{m}(\hat{z}) \equiv m(\hat{z}, \boldsymbol{\Theta}_{\text{fid}}) - \langle \hat{m} \,|\, \mathcal{S}_{\text{o}}, \hat{z}, \boldsymbol{\Theta}_{\text{fid}} \rangle. \tag{C.3}$$

This is also the definition that we employ for figures 4 and 5 and below in this appendix to calculate $\Delta \hat{m}^s \equiv \Delta \hat{m}(\hat{z}^s)$.

### C.2 Linearisation of redshift uncertainties

While proper — Bayesian hierarchical — modelling of significantly uncertain redshift estimates (e.g. coming from photometry) in SN Ia cosmology is computationally expensive, in the absence of selection effects, modern high-dimensional inference techniques can be used to derive a full posterior over the $\mathcal{O}(N_{\text{obs}})$+globals variables. In fact, for our simple model, the latent layer, comprising only the true redshifts, can be numerically marginalised on a grid to evaluate the globals-only marginal likelihood for the purposes of Markov chain Monte Carlo sampling. We used this procedure to derive the cosmological contours from complete samples in the inset of figure 5.

In the presence of selection effects, however, one either needs to run a full BHM analysis like UNITY [24, 25], or subject the data to bias corrections: a non-hierarchical non-Bayesian

---

of selected objects). This may then magnify the systematic bias of inferred parameters resulting from an incorrect bias correction.

procedure, which modifies the observed magnitudes to track the fiducial model without regard for each SN's latent parameters (i.e. the "true" values of noisily measured quantities like redshift, stretch, colour). After bias corrections, inference is performed through a $\chi^2$ fit, which corresponds to assuming Gaussianity for all SN-specific variables, propagating their uncertainties linearly to magnitudes, and combining them in quadrature. For our model, comprising only redshifts and apparent magnitudes, linearisation implies simply

$$\sigma_m^2 \to \sigma_m^2 + \Delta\mu_z^2(\hat{z}, \mathcal{C}) \tag{C.4}$$

with

$$\Delta\mu_z(z, \mathcal{C}) = (1 + \hat{z}^s)\sigma_z \left.\frac{\partial\mu(z, \mathcal{C})}{\partial z}\right|_{z=\hat{z}}. \tag{C.5}$$

In SICRET, we demonstrated that this model delivers biased cosmological results even without selection effects owing to the significant non-linearity in the distance modulus at low redshift and the Eddington bias discussed above, with the discrepancy between true and inferred parameters increasing in significance for larger SN Ia samples. Lastly, we note that this model cannot account for the SN Ia rate $R$ (nor infer its unknown parameters $R_0$, $\beta$), which has been replaced by a (improper) uniform prior on $z$ due to the requirement of analytic marginalisability.

### C.3 Systematic bias from bias correction

Finally, the $\chi^2$ for the "de-biased" model which we use for comparison with STAR NRE in the inset of figure 5 is

$$\chi^2 = \sum_{s=1}^{N_{\rm obs}} \frac{[(\hat{m}^s + \Delta\hat{m}^s) - (M_0 + \mu(\hat{z}^s, \mathcal{C}))]^2}{\sigma_m^2 + \Delta\mu_z^2(\hat{z}^s, \mathcal{C}_{\rm fid})}. \tag{C.6}$$

Note that the redshift-related contribution to the magnitude uncertainty ($\Delta\mu_z$), similarly to the correction $\Delta\hat{m}$, is evaluated at the fiducial cosmology and fixed: we found that calculating different $\Delta\mu_z$ for every sample of $\mathcal{C}$ in a MCMC chain (and adding the usual $\sum_{s=1}^{N_{\rm obs}} \ln[\sigma_m^2 + \Delta\mu_z^2(\hat{z}^s, \mathcal{C})]$ to eq. (C.6)) leads to a significant bias since the fit favours cosmologies with more slowly varying distance modulus, which translates into smaller total uncertainty. This effect may be mitigated by the uncertainty rescaling applied by more involved bias correction and error propagation pipelines [see e.g. 101, subsection 3.5], although we have not tested this.

Regardless, while this linearised bias-corrected model works — i.e. gives consistent contours — when applied to data generated from the fiducial parameters with which is has been constructed (see the inset of figure 5), it fails when the "true" parameters are sufficiently different. As we illustrate in figure 11, for the realistic redshift evolution of the SN Ia rate that we assume (see subsection 3.1) — which dictates that a significant fraction of the total constraining power of the sample is contributed by the high-redshift objects affected by selection effects — the bias correction procedure ends up "imprinting" the fiducial cosmological model, and so inference from any data is compelled to return results in accordance with the fiducial cosmology: this was also noted by [11, figure 4].

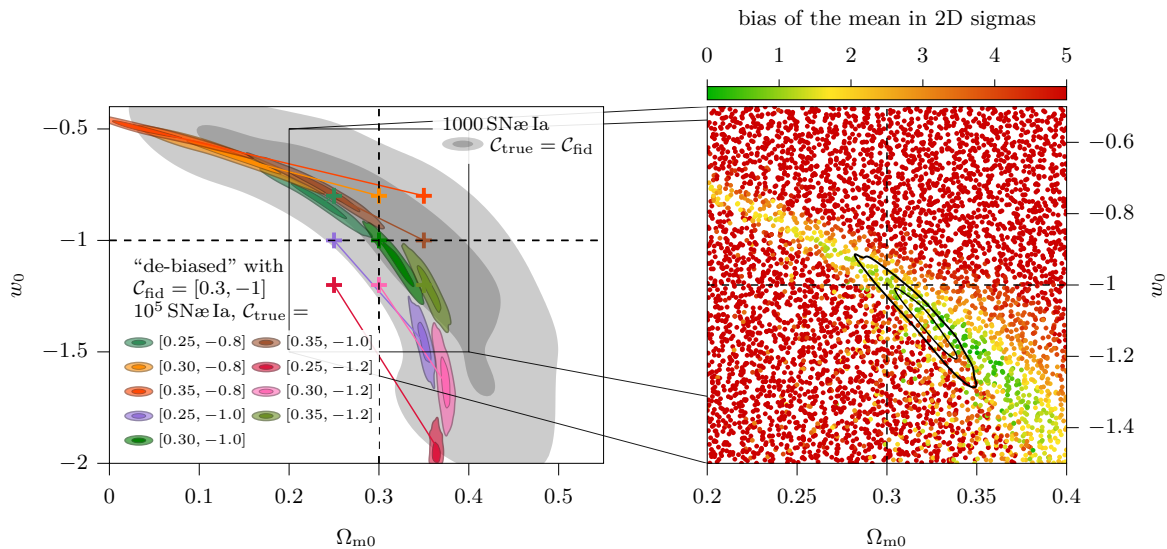

**Figure 11.** Illustration of the systematic bias introduced by bias correction, whereby posteriors (from a likelihood-based analysis using eq. (C.6)) trace the region in $\mathcal{C}$ space most consistent with the fiducial cosmology, $\mathcal{C}_{\mathrm{fid}} = [0.3, -1]$, instead of recovering the true parameters. This characteristic "banana"-shaped region is illustrated in pale grey through the 1- and 2-sigma credible intervals from a bias-corrected analysis of 1000 mock SNæ Ia generated from the fiducial model. *Left:* posteriors (1- and 2-sigma credible regions) from $10^5$ mock SNæ Ia generated with different cosmological parameters $\mathcal{C}_{\mathrm{true}}$ indicated by crosses (and all other global parameters as in table 1). *Right:* magnitude of the systematic bias in the posterior mean in units of the statistical uncertainty (calculated from the posterior covariance matrix) as a function of the true underlying cosmology. For each point, a different mock data set of $10^5$ SNæ Ia was generated and bias-corrected assuming the same $\mathcal{C}_{\mathrm{fid}}$. The black contours replicate the case $\mathcal{C}_{\mathrm{true}} = \mathcal{C}_{\mathrm{fid}}$ from the left panel. The systematic bias severely increases as one moves away from the locus of degeneracy.

This systematic effect is independent of the data, and thus its significance increases with the size and constraining power of the analysed sample. As a demonstration, in figure 11, we test data sets of $\approx 10^5$ SNæ Ia generated with $\Omega_{\mathrm{m}0}$ offset by up to $\pm 0.05$ and $w_0$ by up to $\pm 0.2$ from the fiducial $\mathcal{C}_{\mathrm{fid}} = [0.3, -1]$ and find that all constraints resulting from "de-biased" $\chi^2$ fits, instead of recovering the underlying true values, lie within the characteristic banana-shaped region of parameters most consistent with $\mathcal{C}_{\mathrm{fid}}$, as revealed by an analysis of a smaller mock data set generated with $\mathcal{C}_{\mathrm{fid}}$. Since the constraints from the latter — and similarly from real past and present surveys — are not particularly strong, the systematic bias in the cosmological inference introduced by bias correction is not particularly pronounced for these "small" data sets. But as the SN Ia sample size grows in the future, so will the similarity requirements between the fiducial and true cosmological parameters become more stringent, which means that the validity of bias corrections will rely, paradoxically, on sufficiently precise knowledge of the very parameters whose inference they are supposed to de-bias.

## References

[1] M.M. Phillips, *The absolute magnitudes of type IA supernovae*, *Astrophys. J. Lett.* **413** (1993) L105 [INSPIRE].

[2] R. Tripp, *Using distant type IA supernovae to measure the cosmological expansion parameters*, *Astron. Astrophys.* **325** (1997) 871.

[3] R. Tripp, *A two-parameter luminosity correction for type IA supernovae*, *Astron. Astrophys.* **331** (1998) 815.

[4] K.G. Malmquist, *On some relations in stellar statistics*, *Meddelanden fran Lunds Astronomiska Obs. Ser. I* **100** (1922) 1.

[5] K.G. Malmquist, *A contribution to the problem of determining the distribution in space of the stars*, *Meddelanden fran Lunds Astronomiska Obs. Ser. I* **106** (1925) 1.

[6] SNLS collaboration, *SALT: a Spectral Adaptive Light curve Template for type Ia supernovae*, *Astron. Astrophys.* **443** (2005) 781 [astro-ph/0506583] [INSPIRE].

[7] SNLS collaboration, *SALT2: using distant supernovae to improve the use of type Ia supernovae as distance indicators*, *Astron. Astrophys.* **466** (2007) 11 [astro-ph/0701828] [INSPIRE].

[8] SDSS collaboration, *Improved cosmological constraints from a joint analysis of the SDSS-II and SNLS supernova samples*, *Astron. Astrophys.* **568** (2014) A22 [arXiv:1401.4064] [INSPIRE].

[9] W.D. Kenworthy et al., *SALT3: an improved type Ia supernova model for measuring cosmic distances*, *Astrophys. J.* **923** (2021) 265 [arXiv:2104.07795] [INSPIRE].

[10] A. Möller and T. de Boissière, *SuperNNova: an open-source framework for Bayesian, neural network-based supernova classification*, *Mon. Not. Roy. Astron. Soc.* **491** (2019) 4277.

[11] R. Kessler and D. Scolnic, *Correcting type Ia supernova distances for selection biases and contamination in photometrically identified samples*, *Astrophys. J.* **836** (2017) 56 [arXiv:1610.04677] [INSPIRE].

[12] B. Popovic et al., *Improved treatment of host-galaxy correlations in cosmological analyses with type Ia supernovae*, *Astrophys. J.* **913** (2021) 49 [arXiv:2102.01776] [INSPIRE].

[13] P. Armstrong et al., *Probing the consistency of cosmological contours for supernova cosmology*, *Publ. Astron. Soc. Austral.* **40** (2023) e038 [arXiv:2307.13862] [INSPIRE].

[14] K. Karchev, R. Trotta and C. Weniger, *SICRET: Supernova Ia Cosmology with truncated marginal neural Ratio EsTimation*, *Mon. Not. Roy. Astron. Soc.* **520** (2023) 1056 [arXiv:2209.06733] [INSPIRE].

[15] G. Efstathiou, *Evolving dark energy or supernovae systematics?*, *Mon. Not. Roy. Astron. Soc.* **538** (2025) 875 [arXiv:2408.07175] [INSPIRE].

[16] DES collaboration, *Comparing the DES-SN5YR and Pantheon+ SN cosmology analyses: investigation based on "evolving dark energy or supernovae systematics?"*, arXiv:2501.06664 [INSPIRE].

[17] B. Leistedt, D.J. Mortlock and H.V. Peiris, *Hierarchical Bayesian inference of galaxy redshift distributions from photometric surveys*, *Mon. Not. Roy. Astron. Soc.* **460** (2016) 4258 [arXiv:1602.05960] [INSPIRE].

[18] M. Autenrieth et al., *Improved weak lensing photometric redshift calibration via StratLearn and hierarchical modelling*, *Mon. Not. Roy. Astron. Soc.* **534** (2024) 3808 [arXiv:2401.04687] [INSPIRE].

[19] A.S. Eddington, *On a formula for correcting statistics for the effects of a known probable error of observation*, *Mon. Not. Roy. Astron. Soc.* **73** (1913) 359.

[20] V. Ruhlmann-Kleider, C. Lidman and A. Möller, *Type Ia supernova Hubble diagrams with host galaxy photometric redshifts*, *JCAP* **10** (2022) 065 [arXiv:2207.03789] [INSPIRE].

[21] M.C. March et al., *Improved constraints on cosmological parameters from SNIa data*, *Mon. Not. Roy. Astron. Soc.* **418** (2011) 2308 [arXiv:1102.3237] [INSPIRE].

[22] C. Ma, P.-S. Corasaniti and B.A. Bassett, *Application of Bayesian graphs to SN Ia data analysis and compression*, *Mon. Not. Roy. Astron. Soc.* **463** (2016) 1651 [arXiv:1603.08519] [INSPIRE].

[23] H. Shariff, X. Jiao, R. Trotta and D.A. van Dyk, *BAHAMAS: new analysis of type Ia supernovae reveals inconsistencies with standard cosmology*, *Astrophys. J.* **827** (2016) 1 [arXiv:1510.05954] [INSPIRE].

[24] SUPERNOVA COSMOLOGY PROJECT collaboration, *Unity: confronting supernova cosmology's statistical and systematic uncertainties in a unified Bayesian framework*, *Astrophys. J.* **813** (2015) 137 [arXiv:1507.01602] [INSPIRE].

[25] D. Rubin et al., *Union through UNITY: cosmology with 2,000 SNe using a unified Bayesian framework*, arXiv:2311.12098 [INSPIRE].

[26] DES collaboration, *Steve: a hierarchical Bayesian model for supernova cosmology*, *Astrophys. J.* **876** (2019) 15 [arXiv:1811.02381] [INSPIRE].

[27] M.C. March et al., *A Bayesian approach to truncated data sets: an application to Malmquist bias in supernova cosmology*, arXiv:1804.02474 [INSPIRE].

[28] K.S. Mandel, W.M. Wood-Vasey, A.S. Friedman and R.P. Kirshner, *Type Ia supernova light curve inference: hierarchical Bayesian analysis in the near infrared*, *Astrophys. J.* **704** (2009) 629 [arXiv:0908.0536] [INSPIRE].

[29] K.S. Mandel, G. Narayan and R.P. Kirshner, *Type Ia supernova light curve inference: hierarchical models in the optical and near infrared*, *Astrophys. J.* **731** (2011) 120 [arXiv:1011.5910] [INSPIRE].

[30] K.S. Mandel et al., *A hierarchical Bayesian SED model for type Ia supernovae in the optical to near-infrared*, *Mon. Not. Roy. Astron. Soc.* **510** (2022) 3939 [arXiv:2008.07538] [INSPIRE].

[31] M. Grayling et al., *Scalable hierarchical BayeSN inference: investigating dependence of SN Ia host galaxy dust properties on stellar mass and redshift*, *Mon. Not. Roy. Astron. Soc.* **531** (2024) 953 [arXiv:2401.08755] [INSPIRE].

[32] A.S.M. Uzsoy, S. Thorp, M. Grayling and K.S. Mandel, *Variational inference for acceleration of SN Ia photometric distance estimation with BayeSN*, *Mon. Not. Roy. Astron. Soc.* **535** (2024) 2306 [arXiv:2405.06013] [INSPIRE].

[33] J.M. Szalai-Gindl et al., *GPU-accelerated hierarchical Bayesian estimation of luminosity functions using flux-limited observations with photometric noise*, *Astron. Comput.* **25** (2018) 247.

[34] A. Cole et al., *Fast and credible likelihood-free cosmology with truncated marginal neural ratio estimation*, *JCAP* **09** (2022) 004 [arXiv:2111.08030] [INSPIRE].

[35] K. Cranmer, J. Brehmer and G. Louppe, *The frontier of simulation-based inference*, *Proc. Nat. Acad. Sci.* **117** (2020) 30055 [arXiv:1911.01429] [INSPIRE].

[36] J.-M. Lueckmann, J. Boelts, D. Greenberg, P. Goncalves and J. Macke, *Benchmarking simulation-based inference*, in *Proceedings of the 24th international conference on artificial intelligence and statistics*, https://proceedings.mlr.press/v130/lueckmann21a.html, PMLR, March 2021, p. 343.

[37] A. Weyant, C. Schafer and W.M. Wood-Vasey, *Likelihood-free cosmological inference with type Ia supernovae: approximate Bayesian computation for a complete treatment of uncertainty*, *Astrophys. J.* **764** (2013) 116 [arXiv:1206.2563] [INSPIRE].

[38] E. Jennings, R. Wolf and M. Sako, *A new approach for obtaining cosmological constraints from type Ia supernovae using approximate Bayesian computation*, arXiv:1611.03087 [INSPIRE].

[39] R.C. Bernardo, D. Grandón, J. Levi Said and V.H. Cárdenas, *Dark energy by natural evolution: constraining dark energy using approximate Bayesian computation*, *Phys. Dark Univ.* **40** (2023) 101213 [arXiv:2211.05482] [INSPIRE].

[40] J. Alsing, T. Charnock, S. Feeney and B. Wandelt, *Fast likelihood-free cosmology with neural density estimators and active learning*, *Mon. Not. Roy. Astron. Soc.* **488** (2019) 4440 [arXiv:1903.00007] [INSPIRE].

[41] J. Alsing and B. Wandelt, *Nuisance hardened data compression for fast likelihood-free inference*, *Mon. Not. Roy. Astron. Soc.* **488** (2019) 5093 [arXiv:1903.01473] [INSPIRE].

[42] G.-J. Wang, C. Cheng, Y.-Z. Ma and J.-Q. Xia, *Likelihood-free inference with the mixture density network*, *Astrophys. J. Supp.* **262** (2022) 24 [arXiv:2207.00185] [INSPIRE].

[43] G.-J. Wang et al., *CoLFI: Cosmological Likelihood-Free Inference with neural density estimators*, *Astrophys. J. Suppl.* **268** (2023) 7 [arXiv:2306.11102] [INSPIRE].

[44] J.-F. Chen et al., *Test of artificial neural networks in likelihood-free cosmological constraints: a comparison of information maximizing neural networks and denoising autoencoder*, *Phys. Rev. D* **107** (2023) 063517 [arXiv:2211.05064] [INSPIRE].

[45] V.A. Villar, *Amortized Bayesian inference for supernovae in the era of the Vera Rubin observatory using normalizing flows*, in the proceedings of the *36th conference on neural information processing systems: workshop on machine learning and the physical sciences*, (2022) [arXiv:2211.04480] [INSPIRE].

[46] H. Qu and M. Sako, *Photo-zSNthesis: converting type Ia supernova lightcurves to redshift estimates via deep learning*, *Astrophys. J.* **954** (2023) 201 [arXiv:2305.11869] [INSPIRE].

[47] K. Karchev et al., *SIDE-real: supernova Ia dust extinction with truncated marginal neural ratio estimation applied to real data*, *Mon. Not. Roy. Astron. Soc.* **530** (2024) 3881 [arXiv:2403.07871] [INSPIRE].

[48] K. Karchev, R. Trotta and C. Weniger, *SimSIMS: simulation-based supernova Ia model selection with thousands of latent variables*, in the proceedings of the *37th conference on neural information processing systems*, (2023) [arXiv:2311.15650] [INSPIRE].

[49] T. Allam and J.D. McEwen, *Paying attention to astronomical transients: introducing the time-series transformer for photometric classification*, *RAS Tech. Instrum.* **3** (2023) 209.

[50] C. Donoso-Oliva et al., *ASTROMER: a transformer-based embedding for the representation of light curves*, *Astron. Astrophys.* **670** (2023) A54.

[51] D. Moreno-Cartagena et al., *Positional encodings for light curve transformers: playing with positions and attention*, arXiv:2308.06404.

[52] G. Zhang et al., *Maven: a multimodal foundation model for supernova science*, *Mach. Learn. Sci. Tech.* **5** (2024) 045069 [arXiv:2408.16829] [INSPIRE].

[53] K. Boone, *Avocado: photometric classification of astronomical transients with Gaussian process augmentation*, *Astron. J.* **158** (2019) 257.

[54] P.L.C. Rodrigues, T. Moreau, G. Louppe and A. Gramfort, *HNPE: leveraging global parameters for neural posterior estimation*, in *Advances in neural information processing systems*, volume 34, Curran Associates Inc. (2021), p. 13432 [arXiv:2102.06477].

[55] S.T. Radev et al., *BayesFlow: learning complex stochastic models with invertible neural networks*, *IEEE Trans. Neural Networks Learn. Syst.* **33** (2022) 1452.

[56] È. Campeau-Poirier, L. Perreault-Levasseur, A. Coogan and Y. Hezaveh, *Time delay cosmography with a neural ratio estimator*, in the proceedings of the *40th international conference on machine learning*, (2023) [arXiv:2309.16063] [INSPIRE].

[57] L. Heinrich, S. Mishra-Sharma, C. Pollard and P. Windischhofer, *Hierarchical neural simulation-based inference over event ensembles*, arXiv:2306.12584 [INSPIRE].

[58] M. Zaheer et al., *Deep sets*, in *Advances in neural information processing systems*, volume 30, Curran Associates Inc. (2017).

[59] J. Lee et al., *Set transformer: a framework for attention-based permutation-invariant neural networks*, in *Proceedings of the 36th international conference on machine learning*, PMLR, May 2019, p. 3744.

[60] S. Wagner-Carena et al., *From images to dark matter: end-to-end inference of substructure from hundreds of strong gravitational lenses*, *Astrophys. J.* **942** (2023) 75 [arXiv:2203.00690] [INSPIRE].

[61] T. Geffner, G. Papamakarios and A. Mnih, *Compositional score modeling for simulation-based inference*, arXiv:2209.14249.

[62] T.L. Makinen, J. Alsing and B.D. Wandelt, *Fishnets: information-optimal, scalable aggregation for sets and graphs*, arXiv:2310.03812.

[63] N. Anau Montel and C. Weniger, *Detection is truncation: studying source populations with truncated marginal neural ratio estimation*, in the proceedings of the *36th conference on neural information processing systems: workshop on machine learning and the physical sciences*, (2022) [arXiv:2211.04291] [INSPIRE].

[64] F. Gerardi, S.M. Feeney and J. Alsing, *Unbiased likelihood-free inference of the Hubble constant from light standard sirens*, *Phys. Rev. D* **104** (2021) 083531 [arXiv:2104.02728] [INSPIRE].

[65] S. Gagnon-Hartman, J. Ruan and D. Haggard, *Debiasing standard siren inference of the Hubble constant with marginal neural ratio estimation*, *Mon. Not. Roy. Astron. Soc.* **520** (2023) 1 [arXiv:2301.05241] [INSPIRE].

[66] B.M. Boyd, M. Grayling, S. Thorp and K.S. Mandel, *Accounting for selection effects in supernova cosmology with simulation-based inference and hierarchical Bayesian modelling*, in the proceedings of the *41st international conference on machine learning*, (2024) [arXiv:2407.15923] [INSPIRE].

[67] S.N. Chiu, D. Stoyan, W.S. Kendall and J. Mecke, *Point processes I — The Poisson point process*, in *Stochastic geometry and its applications*, Wiley, U.S.A. (2013).

[68] J. Hermans, V. Begy and G. Louppe, *Likelihood-free MCMC with amortized approximate ratio estimators*, in *Proceedings of the 37th international conference on machine learning*, ICML'20, JMLR.org, July 2020, p. 4239 [`arXiv:1903.04057`] [InSPIRE].

[69] B.K. Miller, A. Cole, P. Forré, G. Louppe and C. Weniger, *Truncated marginal neural ratio estimation*, in *Advances in neural information processing systems*, volume 34, Curran Associates Inc. (2021), p. 129.

[70] N. Anau Montel, J. Alvey and C. Weniger, *Scalable inference with autoregressive neural ratio estimation*, *Mon. Not. Roy. Astron. Soc.* **530** (2024) 4107 [`arXiv:2308.08597`] [InSPIRE].

[71] B. de Finetti, *La prévision: ses lois logiques, ses sources subjectives* (in French), *Ann. Inst. Henri Poincaré* **7** (1937) 1.

[72] D.W. Hogg, *Distance measures in cosmology*, `astro-ph/9905116` [InSPIRE].

[73] SDSS collaboration, *A measurement of the rate of type-Ia supernovae at redshift $z \approx 0.1$ from the first season of the SDSS-II supernova survey*, *Astrophys. J.* **682** (2008) 262 [`arXiv:0801.3297`] [InSPIRE].

[74] LSST Dark Energy Science et al. collaborations, *Models and simulations for the Photometric LSST Astronomical Time series Classification Challenge (PLAsTiCC)*, *Publ. Astron. Soc. Pac.* **131** (2019) 094501 [`arXiv:1903.11756`] [InSPIRE].

[75] R. Hounsell et al., *Simulations of the WFIRST supernova survey and forecasts of cosmological constraints*, *Astrophys. J.* **867** (2018) 23 [`arXiv:1702.01747`] [InSPIRE].

[76] S.A. Rodney et al., *Type Ia supernova rate measurements to redshift 2.5 from CANDELS: searching for prompt explosions in the early universe*, *Astron. J.* **148** (2014) 13 [`arXiv:1401.7978`] [InSPIRE].

[77] O. Graur et al., *Type-Ia supernova rates to redshift 2.4 from CLASH: the Cluster Lensing And Supernova survey with Hubble*, *Astrophys. J.* **783** (2014) 28 [`arXiv:1310.3495`] [InSPIRE].

[78] F. Delgado et al., *The LSST operations simulator*, in the proceedings of the *Modeling, systems engineering, and project management for astronomy VI*, (2014) [`DOI:10.1117/12.2056898`].

[79] P. Yoachim et al., *An optical to IR sky brightness model for the LSST*, in the proceedings of the *Observatory operations: strategies, processes, and systems VI*, (2016) [`DOI:10.1117/12.2232947`].

[80] P. Yoachim et al., *Lsst/rubin_sim: v2.2.4*, Zenodo, November 2025.

[81] LSST Science and LSST Project collaborations, *LSST science book, version 2.0*, `arXiv:0912.0201` [InSPIRE].

[82] J.L. Ba, J.R. Kiros and G.E. Hinton, *Layer normalization*, `arXiv:1607.06450` [InSPIRE].

[83] D.P. Kingma and J. Ba, *Adam: a method for stochastic optimization*, `arXiv:1412.6980` [InSPIRE].

[84] E. Heringer et al., *Type Ia supernovae: colors, rates, and progenitors*, *Astrophys. J.* **834** (2017) 15 [`arXiv:1611.01162`] [InSPIRE].

[85] DES collaboration, *First cosmology results using type Ia supernovae from the Dark Energy Survey: the effect of host galaxy properties on supernova luminosity*, *Mon. Not. Roy. Astron. Soc.* **494** (2020) 4426 [`arXiv:2001.11294`] [InSPIRE].

[86] R. Kessler et al., *SNANA: a public software package for supernova analysis*, *Publ. Astron. Soc. Pac.* **121** (2009) 1028 [`arXiv:0908.4280`] [InSPIRE].

[87] K. Barbary et al., *SNCosmo*, Zenodo, November 2025.

[88] E. Roberts et al., *zBEAMS: a unified solution for supernova cosmology with redshift uncertainties*, *JCAP* **10** (2017) 036 [arXiv:1704.07830] [INSPIRE].

[89] DES collaboration, *The Dark Energy Survey: cosmology results with ∼ 1500 new high-redshift type Ia supernovae using the full 5 yr data set*, *Astrophys. J. Lett.* **973** (2024) L14 [arXiv:2401.02929] [INSPIRE].

[90] C. Li et al., *Multimodal foundation models: from specialists to general-purpose assistants*, arXiv:2309.10020.

[91] A. Paszke et al., *PyTorch: an imperative style, high-performance deep learning library*, in *Advances in neural information processing systems*, H. Wallach et al. eds., Curran Associates Inc. (2019), p. 8024 [arXiv:1912.01703] [INSPIRE].

[92] A. Iana, G. Glavaš and H. Paulheim, *NewsRecLib: a PyTorch-lightning library for neural news recommendation*, arXiv:2310.01146.

[93] D. Foreman-Mackey, D.W. Hogg, D. Lang and J. Goodman, *emcee: the MCMC hammer*, *Publ. Astron. Soc. Pac.* **125** (2013) 306 [arXiv:1202.3665] [INSPIRE].

[94] P. Virtanen et al., *SciPy 1.0 — fundamental algorithms for scientific computing in python*, *Nature Meth.* **17** (2020) 261 [arXiv:1907.10121] [INSPIRE].

[95] J.S. Speagle, *dynesty: a dynamic nested sampling package for estimating Bayesian posteriors and evidences*, *Mon. Not. Roy. Astron. Soc.* **493** (2020) 3132 [arXiv:1904.02180] [INSPIRE].

[96] A. Vehtari et al., *Expectation propagation as a way of life: a framework for Bayesian inference on partitioned data*, arXiv:1412.4869.

[97] S. Srivastava, V. Cevher, Q. Dinh and D. Dunson, *WASP: scalable Bayes via barycenters of subset posteriors*, in *Proceedings of the eighteenth international conference on artificial intelligence and statistics*, PMLR, February 2015, p. 912.

[98] S.L. Scott et al., *Bayes and big data: the consensus Monte Carlo algorithm*, *Int. J. Management Sci. Eng. Manag.* **11** (2016) 78.

[99] C. Vyner, C. Nemeth and C. Sherlock, *SwISS: a scalable Markov chain Monte Carlo divide-and-conquer strategy*, arXiv:2208.04080.

[100] DES collaboration, *Evaluating cosmological biases using photometric redshifts for type Ia Supernova cosmology with the Dark Energy Survey supernova program*, *Mon. Not. Roy. Astron. Soc.* **536** (2024) 1948 [arXiv:2407.16744] [INSPIRE].

[101] DES collaboration, *The Dark Energy Survey supernova program: cosmological analysis and systematic uncertainties*, *Astrophys. J.* **975** (2024) 86 [arXiv:2401.02945] [INSPIRE].